%% file: main.tex
\documentclass{article}

\usepackage{arxiv}
\usepackage[utf8]{inputenc}
\usepackage[T1]{fontenc}
\usepackage{hyperref}
\usepackage{url}
\usepackage{booktabs}
\usepackage{amsmath,amssymb,amsfonts}
\usepackage{graphicx}
\usepackage{microtype}
\usepackage[numbers,sort&compress]{natbib}
\usepackage{cleveref}
\usepackage{enumitem}
\usepackage{optidef}
\usepackage[ruled,vlined,linesnumbered,noend]{algorithm2e}
\usepackage[acronym]{glossaries}
\usepackage{authblk}

\setlist[itemize]{leftmargin=1.5em}
\setlist[enumerate]{leftmargin=1.7em}
\DontPrintSemicolon
\SetAlgoNlRelativeSize{-1}
\SetKwInput{KwInput}{Input}
\SetKwInput{KwOutput}{Output}
\SetKw{KwContinue}{continue}
\SetKw{KwBreak}{break}
\SetKw{KwReturn}{return}

\title{Exoplanetary Tour Design with Solar Sails: TheAntipodes Results in the GTOC13 Problem}
\date{}
\glsdisablehyper

\author[1]{%
	{Jack Yarndley\thanks{Corresponding Author: \texttt{jyar540@aucklanduni.ac.nz}}}%
}
\author[1]{%
	{Adam Evans}%
}
\author[2]{%
	{Xingyu Zhou}%
}
\author[3]{%
	{Minduli Wijayatunga}%
}
\author[1]{%
	{Roberto Armellin}%
}

\affil[1]{Te P\=unaha \=Atea -- Space Institute, University of Auckland, Auckland 1010, New Zealand}
\affil[2]{Beijing Institute of Technology, 100081 Beijing, People's Republic of China}
\affil[3]{University of Illinois Urbana-Champaign, Urbana, IL 61801, United States}

\renewcommand{\headeright}{}
\renewcommand{\undertitle}{}
\renewcommand{\shorttitle}{\textit{Exoplanetary Tour Design with Solar Sails: TheAntipodes Results in the GTOC13 Problem}}

\makeatletter
\renewcommand\maketitle{%
    {{%
        \renewenvironment{tabular}[2][]{%
            \begin{center}\begin{minipage}{1.0\textwidth}\centering
        }{%
            \end{minipage}\end{center}
        }%
        \AB@maketitle
    }}%
}
\makeatother

\hypersetup{
pdftitle={Exoplanetary Tour Design with Solar Sails: TheAntipodes Results in the GTOC13 Problem},
pdfsubject={astro-ph.IM},
pdfauthor={Jack Yarndley, Adam Evans, Xingyu Zhou, Minduli Wijayatunga, Cristina Parigini and Roberto Armellin},
pdfkeywords={GTOC13, solar sailing, gravity assist, beam search, successive convex programming},
}

\input{acronyms.tex}

\begin{document}
\maketitle

\begin{abstract}
Solar sails present an attractive but challenging propulsion method for large-scale, long-duration trajectory design problems. In 2025, the 13th Global Trajectory Optimization Competition (GTOC13) presented a trajectory design problem involving an exoplanetary solar sailing spacecraft in the fictional Altaira system, where the goal is to collect scientific return from flybys of planets, comets, and asteroids. High-scoring solutions combine combinatorial gravity assist tour design with continuous solar sail trajectory optimization. This paper presents the solution approach developed by the team `TheAntipodes' during GTOC13. The approach combines several search and optimization stages: (1) trade studies to identify competitive entry opportunities, (2) large-scale beam search over ballistic gravity assist tours to identify beneficial planetary structures, (3) resonant targeting strategies for Vulcan flyby sequences, and (4) multi-leg solar sail trajectory refinement using sequential convex programming (SCP). A key component of the refinement process is the use of a lossless control-convex solar sail formulation, which allows for large portions of the trajectory, including all gravity assist geometry and flyby timing, to be optimized simultaneously to maximize score. The resulting trajectory placed third, with a score of 337.878 from 133 scoring flybys, and exhibited a structure broadly similar to those of the other high-scoring solutions. This demonstrates the scalability of methods such as SCP for very large trajectory design problems.
\end{abstract}

\glsresetall

\begin{center}
\end{center}

\section{Introduction}

Solar sails harness momentum transfer from photons to provide a propellant-free means of propulsion for spacecraft. Although the resulting acceleration is small when compared with chemical or even electric propulsion systems, it can be applied continuously over very long durations. This makes solar sails attractive for missions where propellant mass is highly constrained, or where large changes in orbital energy and geometry can be obtained through sustained low acceleration \citep{mcinnesSolarSailing1999,berthetSpaceSailsAchieving2024}. Recent solar sail technology developments and demonstrations, including IKAROS, NanoSail-D, NEA Scout, and Solar Cruiser, have further motivated their consideration for practical spacecraft mission design \citep{tsudaAchievementIKAROSJapanese2013,johnsonNanoSailDSolarSail2011,pezentPreliminaryTrajectoryDesign2021,wilkieOverviewNASAAdvanced2021, lantoineTrajectoryManeuverDesign2024}.

From the perspective of trajectory design, the use of solar sailing comes with significant challenges. The acceleration produced by a solar sail is determined by solar radiation pressure, the spacecraft distance from the star, and the orientation of the sail surface relative to the incoming photon direction. Consequently, the achievable acceleration set is strongly constrained: a solar sail cannot produce thrust in an arbitrary direction, and the attainable acceleration magnitude is nonlinearly coupled to the selected direction \citep{mcinnesSolarSailing1999,oguriSolarSailingPrimer2022}. Solar sail spacecraft are therefore highly underactuated dynamical systems and introduce significant nonconvexity into trajectory optimization problems.

The design of solar sail trajectories is part of the broader field of spacecraft trajectory optimization, which includes a wide range of direct, indirect, and heuristic methods \citep{conwaySpacecraftTrajectoryOptimization2010,chaiReviewOptimizationTechniques2019}. Indirect methods exploit necessary conditions of optimality and can produce highly accurate solutions, but are often sensitive to the initial costate guess and can be difficult to extend when additional path constraints, event constraints, or hybrid mission logic are introduced. Direct methods instead discretize the trajectory and solve the resulting nonlinear programming problem, making them generally easier to apply to constrained mission design problems, although the resulting optimization problems are often large, nonconvex, and sensitive to the quality of the initial guess. \Gls{scp} has emerged as a promising direct optimization approach for such problems by approximating the original nonlinear optimal control problem as a sequence of convex subproblems solved around a reference trajectory \citep{maoSuccessiveConvexificationSuperlinearly2019,malyutaConvexOptimizationTrajectory2022}.

The appeal of this approach lies in the efficiency and robustness of convex optimization methods, while still permitting nonlinear dynamics and constraints to be handled through successive linearization, trust regions, and virtual controls. \gls{scp}-based methods have been applied across a range of aerospace problems, including launch vehicle ascent guidance, powered descent guidance, and low-thrust interplanetary trajectory optimization \citep{benedikterConvexApproachThreeDimensional2021,kwonSequentialConvexProgramming2021,hofmannComputationalGuidanceLowThrust2023}. For solar sail trajectory design, it is possible to losslessly convexify the solar sail dynamics, providing a particularly useful mechanism to help solve large-scale direct formulations of solar sail trajectory design problems \citep{oguriLosslessControlConvexFormulation2024}.

In parallel, many modern mission design problems contain a large combinatorial component. Multi-target flyby and rendezvous missions require not only the optimization of the continuous trajectory, but also the selection, sequencing, and timing of target encounters. Such problems are closely related to graph-search and traveling-salesman problem formulations \citep{lawlerTravelingSalesmanProblem1985,conwaySpacecraftTrajectoryOptimization2010}, which can be expressed using \gls{mip} formulations \citep{yarndleyMultitargetSpacecraftMission2026, bannachKeplerianTravelingSalesperson2026}. However, complete enumeration of the search space is typically intractable, and exact mixed-integer approaches can become computationally expensive for large target sets. As a result, practical solution methods often decompose the problem into tractable subproblems, using branch-and-bound, branch-and-cut, dynamic programming, beam search, or stochastic/metaheuristic methods to explore the combinatorial structure \citep{lawlerBranchandBoundMethodsSurvey1966,padbergBranchCutAlgorithmResolution1991,blumMetaheuristicsCombinatorialOptimization2003}. Beam search is particularly attractive in this setting because it follows a breadth-first search strategy while limiting computational effort by retaining only a fixed number of promising partial solutions at each depth. This type of incomplete but scalable search strategy has been used successfully in several large-scale spacecraft trajectory design problems \citep{izzoSearchGrandTour2013, armellinTeamTheAntipodesSolution2022, zhangGlobalTrajectoryOptimization2024, armellinGTOC12ResultsTheAntipodes2025, garciamateasAutomatedPathPlanningAsteroid2026}.

The \gls{gtoc13}, titled ``Humanity's First Robotic Exploration of a Hypothetical Exoplanetary System'', provides a challenging setting in which these ideas must be combined \citep{whiffenGTOC13HumanitysFirst}. The \gls{gtoc} was first proposed by the Advanced Concepts Team at ESA in 2005 \citep{izzo1stACTGlobal2007} and has since become a major benchmark for difficult spacecraft trajectory design problems. Recent editions have increasingly emphasized large-scale, multi-target mission design problems, including the coordinated construction of a Dyson ring in GTOC11 \citep{zhangGTOC11Results2023,armellinTeamTheAntipodesSolution2022} and the design of coordinated asteroid-mining campaigns in GTOC12 \citep{zhangSustainableAsteroidMining2025,armellinGTOC12ResultsTheAntipodes2025}. \gls{gtoc13} continues this trend, but introduces difficulty by requiring the design of a multiple-\gls{ga} trajectory for a spacecraft that only has a solar sail for control through the fictional Altaira star system. A successful solution therefore requires both broad exploration of the combinatorial search space of the possible targets and accurate refinement of the continuous solar sail trajectory.

A strong coupling between these two aspects makes the full \gls{gtoc13} problem computationally impractical to solve as a single optimization problem. This paper presents the approach used by the third-place team, \textit{TheAntipodes}, in which the design is decomposed into a sequence of tractable search, refinement, and assembly stages. First, entry opportunities into the Altaira system are analyzed to identify favorable initial conditions and early \gls{ga} structures. A large-scale ballistic beam search is then used to identify promising planetary flyby sequences, before solar sail refinement is applied to improve timing, flyby geometry, and scientific return. A separate Vulcan-resonant ballistic search is then used to develop tours based on repeated Vulcan encounters. Additional solar sail tour-design stages are used to analyze entry options, refine the ballistic and resonant structures, and assess asteroid-belt flyby tours. The best resulting phases are joined, refined simultaneously, and post-processed to select the scientific flybys used in the submitted trajectory.

The paper is organized as follows. Section~\ref{sec:problem_statement} introduces the \gls{gtoc13} problem statement, scoring model, and ideal flat-plate solar sail model. Section~\ref{sec:ballistic_design} presents the construction of ballistic \gls{ga} sequences, including planetary beam searches and Vulcan-resonant tour searches. Section~\ref{sec:solar_sail_design} describes the \gls{scp} formulation used for trajectory refinement and applies it to ballistic tour refinement, entry-option analysis, terminal-sequence refinement, asteroid-belt tours, and Vulcan-resonant tour refinement. Section~\ref{sec:assembly} explains how the solution phases are joined and how scientific flybys are selected. Finally, Section~\ref{sec:results} presents the submitted trajectory and key observations from the campaign.

\pagebreak

\section{Problem Statement} \label{sec:problem_statement}

\begin{figure}[t]
\centering
\includegraphics[width=\textwidth]{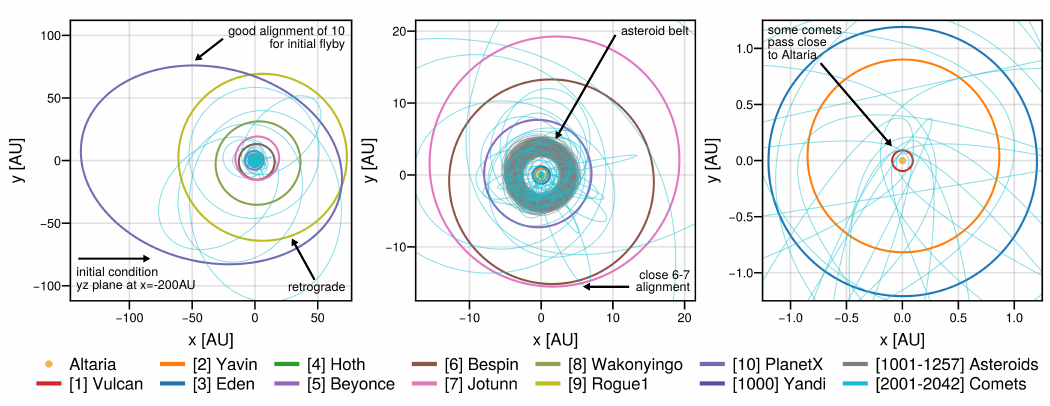}
\caption{Geometry of the Altaira system at several different scales.}
\label{fig:system_overview}
\end{figure}

\begin{table}[t]
\centering
\caption{Parameters of the objects within the Altaira system.}
\begin{tabular}{rlrrrrr}
\toprule
ID & Name & $a$ (AU) & $T$ (yr) & $\mu$ (km$^3$/s$^2$) & Radius (km) & Weight \\
\midrule
1   & Vulcan       & 0.092  & 0.027   & $6.589\cdot 10^8$  & 133021 & 0.1  \\
2   & Yavin        & 0.859  & 0.777   & $6.363\cdot 10^6$  & 18013  & 1    \\
3   & Eden         & 1.200  & 1.283   & $4.439\cdot 10^5$  & 6697   & 2    \\
4   & Hoth         & 2.939  & 4.916   & $2.844\cdot 10^5$  & 5499   & 3    \\
5   & Beyonce      & 7.481  & 19.968  & $4.932\cdot 10^7$  & 63476  & 7    \\
6   & Bespin       & 14.295 & 52.748  & $1.204\cdot 10^8$  & 63661  & 10   \\
7   & Jotunn       & 17.529 & 71.623  & $6.342\cdot 10^6$  & 23865  & 15   \\
8   & Wakonyingo   & 34.195 & 195.145 & $6.598\cdot 10^6$  & 13531  & 20   \\
9   & Rogue1       & 67.173 & 537.289 & $6.635\cdot 10^7$  & 109471 & 35   \\
10  & PlanetX      & 106.653& 1074.908& $3.412\cdot 10^6$ & 12994  & 50   \\
\addlinespace
1000 & Yandi       & 3.986  & 7.767   & ---                & ---    & 5    \\
1001--1257 & Asteroid & 3.1--4.6 & 5.4--9.6 & --- & --- & 1 \\
2001--2042 & Comet    & 2.7--77  & 4.3--662 & --- & --- & 3 \\
\bottomrule
\end{tabular}
\label{tab:weights}
\end{table}

The \gls{gtoc13} problem considers the exploration of the fictional Altaira exoplanetary system with a spacecraft equipped only with a solar sail for maneuvering \citep{whiffenGTOC13HumanitysFirst}. The problem requires the design of a spacecraft tour through the Altaira system that begins on an interstellar approach and subsequently visits a sequence of objects within the system to maximize scientific return from flybys of planets, asteroids, comets, and a dwarf planet.

The Altaira system, shown at several scales in Fig.~\ref{fig:system_overview}, contains 10 major planets (1-10), the dwarf planet Yandi (1000), 257 main-belt asteroids (1001-1257), and 42 comets (2001-2042). The main-belt asteroids lie between the orbits of Hoth (4) and Beyonce (5), while the comets are distributed more broadly throughout the system. The major planets, asteroids, and comets follow Keplerian orbits about Altaira, with the orbital plane of Vulcan (1) defining the ecliptic of the system. The object parameters and raw scientific weights are summarized in Table~\ref{tab:weights}.

\begin{figure}[t]
\centering
\includegraphics[width=3.25in]{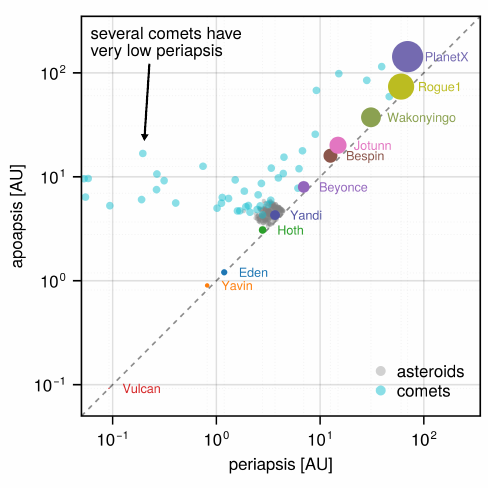}
\caption{Comparison between object periapsis and apoapsis in the Altaira system.}
\label{fig:system_weights}
\end{figure}

Importantly, only the 10 major planets can be used for \gls{ga}s since they have gravitational parameters assigned. The \gls{ga}s are modeled using patched conics, with the incoming and outgoing flyby velocities constrained to have equal magnitude. Each planetary flyby must occur at an altitude between $0.1$ and $100$ body radii above the surface, equivalent to a periapsis radius between $1.1$ and $101$ body radii from the body center. All of the remaining bodies are treated as massless bodies. Flybys of these objects must therefore be dynamically continuous, with no change in the spacecraft velocity relative to the body.

This distinction is critically important for the construction of feasible tours. For massless bodies, a flyby can contribute to the total score but cannot change the trajectory direction. Therefore, without the use of the solar sail, subsequent targeting of another object after such a flyby requires very careful design, unless the encounter already lies on a favorable trajectory. For this reason, the effective use of the solar sail is central to the construction of trajectories that include asteroids, comets, and the dwarf planet Yandi within a broader \gls{ga} tour.

Generally, objects further from Altaira have larger raw scientific weights than those in the inner system. This is clearly apparent in the plot of periapsis and apoapsis of the objects in Fig.~\ref{fig:system_weights}, where the area of the markers is directly proportional to the raw scientific weight of each flyby. These larger weights incentivize strategies that reach the outer parts of the system, at the cost of achieving fewer flybys because of the long transfer times required.

Another important planet to consider is Vulcan, the innermost planet. Vulcan is particularly noteworthy because it has a very large gravitational parameter and short orbital period. This permits large turning angles for \gls{ga}s even for relatively high flyby velocities, which are generally expected so close to Altaira. Therefore, it offers a unique opportunity to redirect the spacecraft trajectory in ways that are not possible with any other planet or with the solar sail. In combination with the low periapsis of several comets, this arrangement underpins one of the core strategies for the \gls{gtoc13} problem developed in this paper: targeting comets through repeated \gls{ga}s at Vulcan.

\begin{table}[t]
\centering
\caption{Problem-specific parameters of GTOC13.}
\begin{tabular}{lc}
\toprule
Parameter & Value \\
\midrule
Altaira gravitational parameter $\mu$ & $1.3935 \cdot 10^{11}~\text{km}^3/\text{s}^2$ \\
Altaira solar flux $C$ at $1~\text{AU}$ & $5.4026 \cdot 10^{-6}~\text{N}/\text{m}^2$ \\
Reference distance $r_0$ & $149597870.691~\text{km}$ \\
Sail area $A$ & $15000~\text{m}^2$ \\
Spacecraft mass $m$ & $500~\text{kg}$ \\
Minimum special periapsis & $0.01~\text{AU}$ \\
Minimum general periapsis & $0.05~\text{AU}$ \\
Trajectory time window & $200~\text{yr}$ \\
\bottomrule
\end{tabular}
\label{tab:constants}
\end{table}

The spacecraft arrives at the Altaira system on an interstellar approach. From this point on, the spacecraft must rely purely on ballistic motion, planetary \gls{ga}s, and optional solar sail maneuvers. A valid trajectory consists of the initial conditions, a time-ordered sequence of flybys, and, when the sail is used, the solar sail control history. All trajectory events must occur within a fixed 200 year time window measured from the reference epoch. The main problem-specific constants are summarized in Table~\ref{tab:constants}.

The coordinate frame is defined such that the $x$ direction is aligned with the incoming asymptotic velocity direction, positive toward Altaira. The $z$ direction is perpendicular to the ecliptic plane and points in the direction of Vulcan's orbital angular momentum, while the $y$ axis completes the right-handed frame. The initial spacecraft position is fixed at $x=-200$ AU, with free $y$ and $z$ components. The initial velocity is constrained to have only an $x$ component, with $v_y=v_z=0$, and the initial epoch $t_0$ may be chosen anywhere in the 200 year trajectory window, although an initial epoch near the start is clearly beneficial because it maximizes the trajectory duration.

The Cartesian equations of motion for the spacecraft are expressed in the ecliptic inertial frame as follows:
\begin{equation}
\dot{\boldsymbol{x}}
=
\begin{bmatrix}
\dot{\boldsymbol{r}} \\
\dot{\boldsymbol{v}}
\end{bmatrix}
=
\begin{bmatrix}
\boldsymbol{v} \\
-\dfrac{\mu}{r^3}\boldsymbol{r}+\boldsymbol{a}_{\mathrm{sail}}
\end{bmatrix}
\label{eq:dynamics-original}
\end{equation}
where $\boldsymbol{r}$ is the spacecraft position vector, $\boldsymbol{v}$ is the spacecraft velocity vector, $\mu$ is the gravitational parameter of Altaira, and $\boldsymbol{a}_{\mathrm{sail}}$ is the acceleration due to the solar sail. The mass of the spacecraft is constant so does not appear in the equations of motion.

\begin{figure}[t]
    \centering
    \includegraphics[width=3.25in]{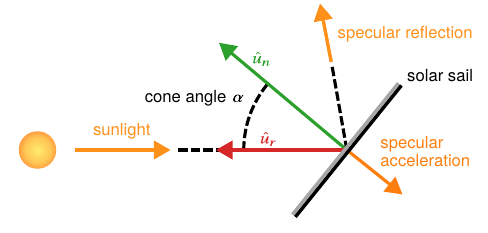}
    \caption{Ideal flat-plate solar sail acceleration model.}
    \label{fig:srp_flat_plate}
\end{figure}

The solar sail may be used over any interval after the starting epoch. A trajectory may alternate between ballistic arcs and solar sail arcs, effectively allowing the solar sail to be deployed or retracted instantaneously. The ideal flat-plate solar sail model \citep{mcinnesSolarSailing1999}, illustrated in Fig.~\ref{fig:srp_flat_plate}, is used to model the effects of \gls{srp}. This model assumes perfect reflection and neglects eclipses. Therefore, when the solar sail is deployed, the corresponding acceleration is expressed by
\begin{equation}
\boldsymbol a_{\mathrm{sail}}
=
-\frac{2CA}{m}
\left(\frac{r_0}{r}\right)^2
(\hat{\boldsymbol u}_n \cdot \hat{\boldsymbol u}_r)^2
        \hat{\boldsymbol u}_n
\label{eq:ideal_sail}
\end{equation}
where $C$ is the Altaira flux at $1$ AU, $A$ is the solar sail area, $m$ is the spacecraft mass, $r_0=1$ AU is the reference distance, $r$ is the spacecraft distance from Altaira, $\hat{\boldsymbol u}_r$ is the unit vector from the spacecraft to Altaira, and $\hat{\boldsymbol u}_n$ is the inward-pointing solar sail normal.

The solar sail has an area-to-mass ratio of $30~\text{m}^2/\text{kg}$, which is approximately double the values associated with several currently proposed or flown solar sail missions \citep{tsudaAchievementIKAROSJapanese2013,johnsonNanoSailDSolarSail2011,pezentPreliminaryTrajectoryDesign2021,wilkieOverviewNASAAdvanced2021,lantoineTrajectoryManeuverDesign2024}. Although the spacecraft begins far from Altaira, the inverse-square dependence of the sail acceleration makes the sail increasingly effective once the trajectory reaches the inner system. This is limited by specific periapsis constraints: all close approaches to Altaira must remain above 0.05 AU, except for a single protected periapsis passage that may reach as low as 0.01 AU.

The competition score is based on a weighted sum of scientific flybys, modified by two global multiplicative bonuses. A maximum of 13 flybys of each body may be considered in the scoring, but more non-scoring flybys are permitted. The grand-tour bonus $b$ gives a $20\%$ score increase to trajectories that visit all 10 major planets, the dwarf planet Yandi, and at least 13 asteroids or comets. The time bonus $c$ depends on the submission date during the competition window, with a maximum value of $13\%$ and a minimum value of $2.5\%$. Since these two bonuses are either fixed by submission time or depend only on a broad coverage condition, they can be considered independent of the local trajectory geometry considered during much of the search and refinement process. The total score is written as
\begin{equation}
J = b\,c \sum_{k \in \mathcal{K}} w_k
\sum_{i=1}^{N_k}
S(\hat{\boldsymbol r}_{k,i}) F(V_{\infty,k,i})
\label{eq:score}
\end{equation}
where $k$ is the body ID, $i$ refers to the $i$th scientific flyby of that body, $w_k$ is the raw scientific weight listed in Table~\ref{tab:weights}, $N_k \leq 13$ is the number of scoring scientific flybys of body $k$, $S$ is the seasonal diversity penalty, and $F$ is the flyby-velocity penalty. Additional flybys of a body are permitted, but only up to 13 scientific flybys per body can contribute to the score.

\begin{figure}[t]
\centering
\includegraphics[width=3.25in]{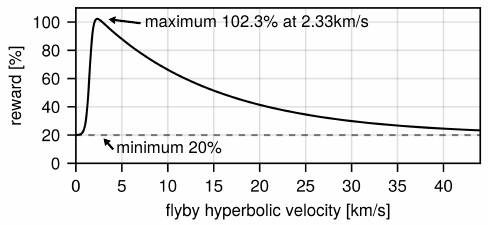}
\caption{Flyby-velocity penalty function.}
\label{fig:velocity_penalty}
\end{figure}

The flyby-velocity penalty, visualized in Fig.~\ref{fig:velocity_penalty}, is defined as
\begin{equation}
F(V_{\infty}) =
0.2 + \frac{\exp\left(-V_{\infty}/13\right)}
{1 + \exp\left(-5(V_{\infty}-1.5)\right)}
\label{eq:velocity_penalty}
\end{equation}
where $V_{\infty}$ is expressed in $\text{km}/\text{s}$. This function penalizes high-speed flybys because they provide shorter observation times, while also penalizing rendezvous-like encounters because of radiation risk and environmental uncertainty near the body. The function has a floor of $20\%$ and can provide a small bonus relative to the nominal body weight, with a maximum value of approximately $102.3\%$ at $V_{\infty}=2.33~\text{km}/\text{s}$.

\begin{figure}[t]
\centering
\includegraphics[width=3.25in]{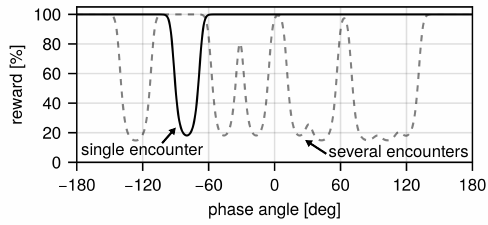}
\caption{Seasonal penalty function for two encounter configurations.}
\label{fig:seasonal_penalty}
\end{figure}

The seasonal diversity penalty, visualized in Fig.~\ref{fig:seasonal_penalty}, penalizes repeated observations of the same body from similar heliocentric viewing directions. It is defined as
\begin{equation}
S(\theta_{k,ij}) =
0.1 +
\frac{0.9}
{1 + 10 \sum_{j=1}^{i-1}
\exp\left(-\theta_{k,ij}^2/50\right)}
\label{eq:seasonal_penalty}
\end{equation}
where $\theta_{k,ij}$ is the angle, in degrees, between the heliocentric unit position vectors of body $k$ at the $i$th and $j$th scientific flybys. The first scientific flyby of each body has $S=100\%$. Subsequent flybys are penalized if they occur at similar solar phase angles, reflecting the reduced scientific value of repeated observations under similar illumination conditions. The seasonal penalty has a floor of $10\%$ and cannot increase the score.

\section{Ballistic Tour Design}
\label{sec:ballistic_design}

The construction of ballistic \gls{ga} tour sequences provided a natural first step in the \gls{gtoc13} solution process. By neglecting the use of the solar sail, the problem reduces to the design of a patched-conic sequence of Keplerian arcs connected by unpowered planetary \glspl{ga}. Such a tour would be expected to be inferior to one that exploits solar sailing, but it provided a useful baseline for the achievable score, helped reveal favorable \gls{ga} structures, helped identify regions of the Altaira system where use of the solar sail was likely to be most valuable, and, most importantly, provided a foundation for the further development of search strategies.

The ballistic interpretation of the problem is more restrictive than may first be expected. Since Yandi, the asteroids, and the comets are modeled as massless bodies, encounters with these objects cannot alter the spacecraft velocity. A ballistic flyby of a massless body is therefore dynamically continuous and is likely to compromise the ability to target subsequent objects, unless the encounter already lies on a favorable transfer. For this reason, the ballistic search has two different approaches: one to find feasible sequences of major-planet \glspl{ga}, and another to identify high-quality \gls{ga} resonant sequences to target the massless bodies.

A beam search was used as the primary tool for identifying ballistic tour sequences for each of these approaches. Its application to the \gls{gtoc13} problem is complicated by two problem-specific features. First, the initial spacecraft state is not fixed, but lies on the incoming plane defined by the interstellar approach condition, so suitable entry conditions must be identified before the planetary sequence can be expanded. Second, the trajectory is limited to a fixed 200-year time window, which requires the beam search metric to balance accumulated scientific value against the time consumed by each partial tour.

Planetary \glspl{ga} are modeled using a patched-conic model. At an encounter with body $b$, the incoming and outgoing spacecraft velocities are converted into the relative flyby velocities using the planet's velocity $\boldsymbol{v}_b$:
\begin{align}
    \boldsymbol{v}_{\infty}^{-} &= \boldsymbol{v}^{-} - \boldsymbol{v}_b \\
    \boldsymbol{v}_{\infty}^{+} &= \boldsymbol{v}^{+} - \boldsymbol{v}_b
    \label{eq:ga-vinf-def}
\end{align}
For an unpowered flyby, the flyby speed is conserved,
\begin{equation}
    \|\boldsymbol{v}_{\infty}^{-}\| = \|\boldsymbol{v}_{\infty}^{+}\| = v_\infty
    \label{eq:ga-speed-match}
\end{equation}
so the \gls{ga} can only rotate the flyby velocity vector. The turning angle $\delta$ can then be calculated:
\begin{equation}
    \delta =
    \arccos\!\left(
    \frac{\boldsymbol{v}_{\infty}^{-\top}\boldsymbol{v}_{\infty}^{+}}
    {\|\boldsymbol{v}_{\infty}^{-}\|\|\boldsymbol{v}_{\infty}^{+}\|}
    \right)
    \label{eq:ga-turning-angle}
\end{equation}
Depending on the limits on the flyby periapsis $r_p$, the achievable turning angles can be calculated based on the flyby speed $v_\infty$ and the gravitational parameter $\mu_b$ of the flyby body:
\begin{equation}
    \delta(r_p,v_\infty) =
    2\arcsin\!\left(\frac{1}{1+r_p v_\infty^2/\mu_b}\right)
    \label{eq:ga-delta-rp}
\end{equation}
Equivalently this can be inverted to find the flyby periapsis $r_p$ in terms of the turning angle $\delta$:
\begin{equation}
    r_p(\delta,v_\infty)
    =
    \frac{\mu_b}{v_\infty^2}
    \left(\frac{1}{\sin(\delta/2)}-1\right)
    \label{eq:ga-rp-from-delta}
\end{equation}
These equations show the trade-off between the flyby periapsis and the maximum deflection: for a fixed $v_\infty$, decreasing the periapsis radius $r_p$ increases the turning angle, while increasing $r_p$ decreases it. The periapsis limits of the \gls{gtoc13} problem therefore constrain the achievable turning angle: the turning angle for a feasible unpowered \gls{ga} must satisfy
\begin{equation}
    \delta(r_{p,\max},v_\infty) \leq \delta \leq \delta(r_{p,\min},v_\infty)
    \label{eq:ga-delta-feasible}
\end{equation}

The geometry of such \glspl{ga} can be represented and visualized using the B-plane \citep{kiznerMethodDescribingMiss1961}. The B-plane is the plane through the flyby body that is perpendicular to the incoming flyby velocity direction
\begin{equation}
    \hat{\boldsymbol{s}} =
    \frac{\boldsymbol{v}_{\infty}^{-}}{\|\boldsymbol{v}_{\infty}^{-}\|}
\end{equation}
For a feasible flyby, the periapsis radius implied by the required turn gives the corresponding magnitude of $\boldsymbol{B}$, the impact parameter:
\begin{equation}
    B = r_p \sqrt{1+\frac{2\mu_b}{r_p v_\infty^2}}
    \label{eq:ga-b-magnitude}
\end{equation}
Then $\boldsymbol{B}$ can be obtained from the flyby plane through a sequence of operations:
\begin{align}
    \hat{\boldsymbol{h}} &=
    \frac{\boldsymbol{v}_{\infty}^{-}\times\boldsymbol{v}_{\infty}^{+}}
    {\|\boldsymbol{v}_{\infty}^{-}\times\boldsymbol{v}_{\infty}^{+}\|}\\
    \hat{\boldsymbol{B}} &= \hat{\boldsymbol{s}}\times\hat{\boldsymbol{h}}\\
    \boldsymbol{B} &= B\hat{\boldsymbol{B}}
    \label{eq:ga-b-vector}
\end{align}
For visualization, an orthonormal basis $(\hat{\boldsymbol{T}},\hat{\boldsymbol{R}})$ is chosen in the B-plane and the aiming point of the \gls{ga} is shown using the coordinates $(B_T,B_R)=(\boldsymbol{B}\cdot\hat{\boldsymbol{T}},\boldsymbol{B}\cdot\hat{\boldsymbol{R}})$. The B-plane is useful for inspecting the geometry of the transfers: nearby points correspond to similar flyby aim points and similar rotations of the incoming flyby velocity.

\subsection{Finding Ballistic Gravity Assists} \label{sec:ballistic_ga}

The core part of a beam search is the generation of the possible next moves from each node. In the major-planet ballistic \gls{ga} problem, a node is defined by the current encounter body, epoch, and incoming flyby velocity. A valid move is then a Lambert transfer to a subsequent major planet for which the required departure flyby velocity can be produced by an unpowered \gls{ga} at the current body. In other words, the transfer must have a matching flyby speed at departure, while the change in direction must remain within the allowable flyby turning angle.

\begin{figure}[t]
    \centering
    \includegraphics[width=\textwidth]{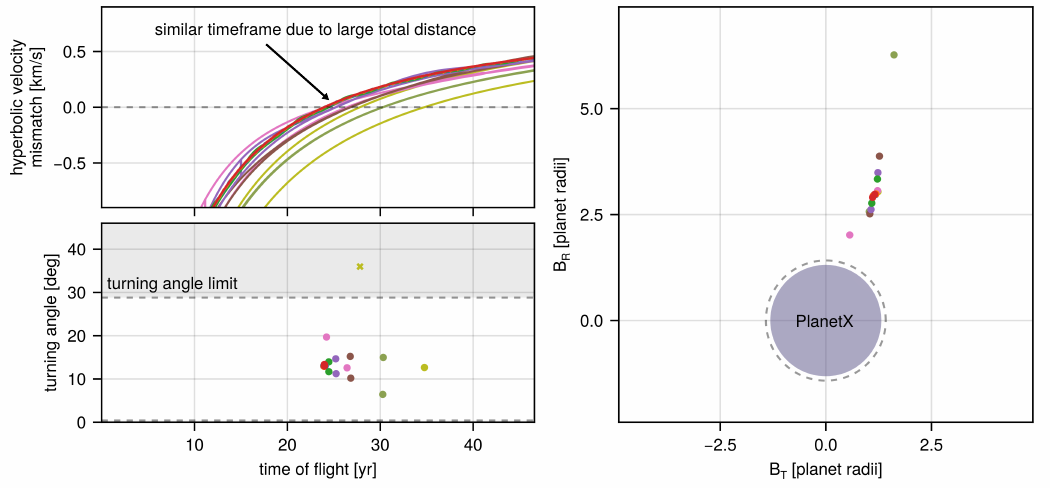}
    \caption{Ballistic GA search geometry starting from PlanetX to all other planets.}
    \label{fig:beam_bplane_10}
\end{figure}

\begin{figure}[t]
    \centering
    \includegraphics[width=\textwidth]{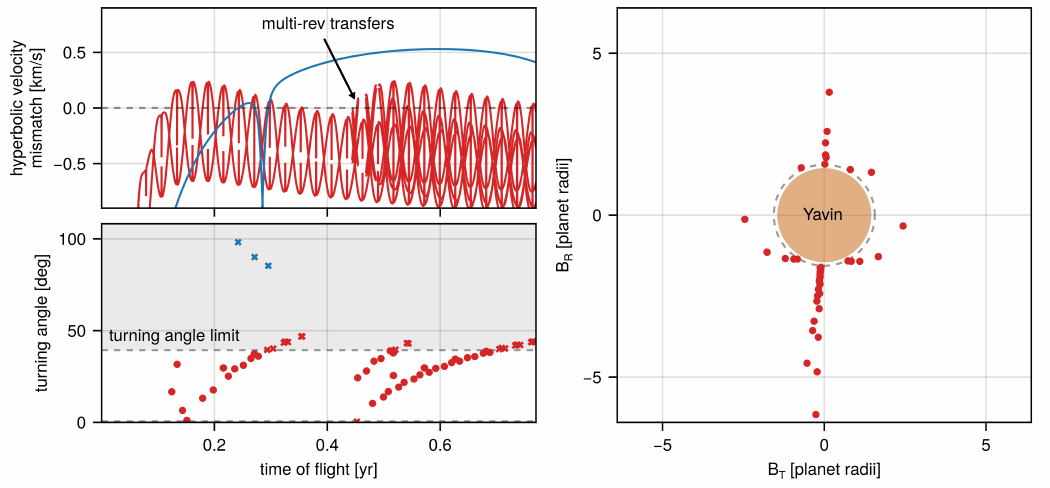}
    \caption{Ballistic GA search geometry starting from Yavin to all other planets.}
    \label{fig:beam_bplane_2}
\end{figure}

Candidate moves are generated by searching over Lambert transfers to each possible next planet. All Lambert geometries are checked, including retrograde and multi-revolution branches. For a fixed departure epoch, target body, and Lambert branch, varying the time of flight changes the required departure flyby velocity. A one-dimensional search over the time of flight can therefore be used to identify ballistic transfers for which the Lambert departure velocity lies on the set of velocities reachable through the current \gls{ga}. This is expressed through the flyby velocity mismatch function
\begin{equation}
    g(\Delta t)=\|\boldsymbol{v}_{\infty}^{+}(\Delta t)\|-\|\boldsymbol{v}_{\infty}^{-}\|
    \label{eq:beam-vinf-mismatch}
\end{equation}
where $\boldsymbol{v}_{\infty}^{+}(\Delta t)$ is the outgoing flyby velocity required for Lambert transfer to the target body with time of flight $\Delta t$.

For each candidate target, the outgoing Lambert transfer time of flight is varied. The signed mismatch between the required outgoing flyby speed and the incoming flyby speed at the current planet is then recorded. When this mismatch is zero, it indicates a possible ballistic \gls{ga}. There are two further conditions on this: the required turning angle must be within the allowable turning angle, and the periapsis of the resulting Keplerian arc must be above the minimum periapsis limit.

This process is illustrated in Fig.~\ref{fig:beam_bplane_10}. Each possible target planet produces a different mismatch curve and, in this case, all produce candidate roots. The turning-angle check then removes a single possible root. The B-plane geometry of all valid \glspl{ga} is then shown. Feasible transfers from PlanetX to the other planets occupy a similar region of the B-plane, which is consistent with transfers from the outermost planet to the inner planets occurring in broadly similar directions.

In some cases, the retrograde, multi-revolution and long-way branches of the Lambert problem become useful to consider. An example of this search is shown in Fig.~\ref{fig:beam_bplane_2}. Here, only transfers to Vulcan and Eden are potentially feasible, and the mismatch curves are much more complex. The development of retrograde and multi-revolution branches is apparent, and the resulting B-plane geometry is more complex than in the previous example. The complexity of the mismatch curves is a result of the short orbital period of Vulcan, which produces rapidly varying transfer geometry. Retaining these opportunities in the beam search was important for improving solution quality by allowing phasing opportunities involving flybys of the inner planets.

\begin{table}[t]
\centering
\caption{Number of brackets used for root finding over the time of flight.}
\begin{tabular}{lr}
\toprule
Planet & Brackets \\
\midrule
Vulcan & 400 \\
Yavin & 200 \\
Eden & 100 \\
Hoth & 80 \\
Beyonce & 50 \\
Bespin & 50 \\
Jotunn & 30 \\
Wakonyingo & 20 \\
Rogue1 & 20 \\
PlanetX & 20 \\
\bottomrule
\end{tabular}
\label{tab:bracket_sizes}
\end{table}

To identify the roots of the flyby-velocity mismatch, a bracketing strategy is used over the Lambert time of flight. For a node at time $t$ and current body $i$, the heuristic for the admissible time interval $\tau$ for a transfer to body $j$ was
\begin{align}
    \tau_{\min} &= 0.01\min(P_i,P_j)\\
    \tau_{\max} &= \min(0.6\max(P_i,P_j),T_{\max}-t)
    \label{eq:beam-tof-heuristic}
\end{align}
where $P_i$ is the orbital period of the \gls{ga} planet, $P_j$ is the orbital period of the target planet, and $T_{\max}$ is the maximum time limit for the \gls{ga} sequence. This heuristic avoids spending search effort on either extremely short transfers that would not satisfy the \gls{ga} condition or excessively long transfers that would consume too much of the mission window, and also removes all cases that exceed the maximum time limit. The resulting interval between $\tau_{\min}$ and $\tau_{\max}$ is then divided uniformly into a fixed number of brackets for each target planet, as listed in Table~\ref{tab:bracket_sizes}, and the flyby velocity mismatch function $g$ is then evaluated at the edges of each bracket. More brackets were assigned to the inner planets because their shorter orbital periods produce more rapidly varying transfer geometry and therefore a larger number of possible roots.

If a sign change is detected within a bracket, a scalar bracketing root-finding method is applied to refine the corresponding time of flight for the ballistic transfer. During the competition, this refinement used the \texttt{ITP} method \citep{oliveiraEnhancementBisectionMethod2021} implemented in \texttt{NonlinearSolve.jl} \citep{pal2024nonlinearsolve}. Later analysis has found that the newer \texttt{modAB} method \citep{ganchovskiImprovementsModifiedAnderson2026} gave slightly better performance in this application. The Izzo method \citep{izzoRevisitingLambertsProblem2015} was used to solve each Lambert problem in this analysis.

\subsection{Beam Search for Ballistic Tours}
\label{sec:beam_search}

The initial conditions for the beam search were generated using the \gls{scp} formulation described in Section~\ref{sec:scp}. The initial constraint is set to the \gls{gtoc13} approach condition, and the terminal constraint is set to the candidate first flyby planet. For each candidate target, a set of fixed first-leg durations is prescribed. The \gls{scp} problem is then solved for each duration with the objective of maximizing the admissible initial velocity. This produces a corresponding set of flyby conditions at the target planet. For example, in the PlanetX campaign, 50 candidate durations spanning approximately 7--10 years are solved to produce 50 corresponding PlanetX flyby conditions. These formed a catalog of root nodes for the beam search.

From this catalog of root nodes, the beam search expands in a breadth-first manner following Algorithm~\ref{alg:ballistic_beam_search}. At a given search depth, the current retained set of nodes is referred to as the frontier, which has a maximum size $K$. Each frontier node stores the current planet, encounter epoch, cumulative time of flight, cumulative score estimate, encounter history, and incoming flyby velocity state. The feasible ballistic moves described in Section~\ref{sec:ballistic_ga} are generated from each frontier node, producing a large set of candidate nodes at the next depth.

\begin{algorithm}[t]
\caption{Ballistic GA tour beam search.}
\label{alg:ballistic_beam_search}
\KwInput{Initial frontier $\mathcal{F}_0$, candidate flyby moves $\mathcal{M}$, frontier size $K$, elite count $K_E$, ranking objective $\Phi$}
\KwOutput{Elite ballistic tour nodes $\mathcal{E}$}
$\mathcal{F}\gets\mathcal{F}_0$, $\mathcal{E}\gets\mathcal{F}_0$\tcp*{$\mathcal{F}$ is the retained frontier and $\mathcal{E}$ stores persistent elites}
\While{$\mathcal{F}\neq\emptyset$}{
    $\mathcal{C}\gets\emptyset$\tcp*{candidates produced by the next expansion}
    \ForEach{node $n=(i,t,T,W,H,\boldsymbol{v}_{\infty}^{-},Q)\in\mathcal{F}$}{
        \ForEach{move $m=(j,b_{\mathrm{retro}},N_{\mathrm{rev}},b_{\mathrm{long}})\in\mathcal{M}$}{
            Compute admissible time-of-flight interval $[\tau_{\min},\tau_{\max}]$\;
            $(\delta_{\min},\delta_{\max})\gets\mathrm{TurningLimits}(\|\boldsymbol{v}_{\infty}^{-}\|,r_{p,i}^{\max},r_{p,i}^{\min},\mu_i)$\tcp*{turning-angle bounds at body $i$}
            Partition $[\tau_{\min},\tau_{\max}]$ into Lambert root brackets $\{[\tau_a,\tau_b]\}$\tcp*{bracket count from Table~\ref{tab:bracket_sizes}}
            \ForEach{bracket $[\tau_a,\tau_b]$}{
                Define $\boldsymbol{v}_{\infty}^{+}(\tau)\gets\mathrm{LambertOut}(\boldsymbol{r}_i(t),\boldsymbol{r}_j(t+\tau),\tau,b_{\mathrm{retro}},N_{\mathrm{rev}},b_{\mathrm{long}})-\boldsymbol{v}_i(t)$\;
                Define $g(\tau)\gets\|\boldsymbol{v}_{\infty}^{+}(\tau)\|-\|\boldsymbol{v}_{\infty}^{-}\|$\tcp*{flyby-velocity mismatch}
                \If{$g(\tau_a)g(\tau_b)\geq0$}{
                    \KwContinue\;
                }
                $\tau^\ast\gets\mathrm{RootSolve}(g(\tau)=0,[\tau_a,\tau_b])$\tcp*{unpowered \gls{ga} speed match}
                $\delta_{\mathrm{req}}\gets\delta(\boldsymbol{v}_{\infty}^{-},\boldsymbol{v}_{\infty}^{+}(\tau^\ast))$\tcp*{required \gls{ga} turn}
                Reject if $\delta_{\mathrm{req}}<\delta_{\min}$ or $\delta_{\mathrm{req}}>\delta_{\max}$\;
                Reject if $\mathrm{Periapsis}(\boldsymbol{r}_i(t),\boldsymbol{v}_i(t)+\boldsymbol{v}_{\infty}^{+}(\tau^\ast),\tau^\ast)\leq r_{\lim}$\tcp*{Altaira periapsis constraint}
                $\boldsymbol{v}_{\infty}^{-}\gets\mathrm{LambertIn}(\boldsymbol{r}_i(t),\boldsymbol{r}_j(t+\tau^\ast),\tau^\ast,b_{\mathrm{retro}},N_{\mathrm{rev}},b_{\mathrm{long}})-\boldsymbol{v}_j(t+\tau^\ast)$\;
                $\Delta W\gets w_jF(\|\boldsymbol{v}_{\infty}^{-}\|)S(\hat{\boldsymbol{r}}_j(t+\tau^\ast),H)$, $W'\gets W+\Delta W$\;
                $H'\gets\mathrm{Append}(H,(j,t+\tau^\ast,\hat{\boldsymbol{r}}_j(t+\tau^\ast)))$\;
                $n_{\mathrm{child}}\gets(j,t+\tau^\ast,T+\tau^\ast,W',H',\boldsymbol{v}_{\infty}^{-})$\;
                $Q'\gets\Phi(n_{\mathrm{child}})$
                Append $(j,t+\tau^\ast,T+\tau^\ast,W',H',\boldsymbol{v}_{\infty}^{-},Q')$ to $\mathcal{C}$\;
            }
        }
    }
    $\mathcal{E}\gets\operatorname{TopK}(\mathcal{E}\cup\mathcal{C},K_E)$ by total unbonused score\tcp*{stores nodes with best score}
    \If{$\mathcal{C}=\emptyset$}{
        \KwBreak\;
    }
    $\mathcal{F}\gets\operatorname{TopK}(\mathcal{C},K)$ by highest ranking objective\tcp*{stores nodes with best objective}
}
\KwReturn $\mathcal{E}$\;
\end{algorithm}

After each expansion, the candidate nodes are ranked before the best $K$ are retained as the next frontier. Several scalar ranking objectives were used during the campaign. The first maximized unbonused score rate,
\begin{equation}
    R_w = \frac{J}{T}
    \label{eq:beam-max-weight-rate}
\end{equation}
where $J$ is the accumulated unbonused score estimate, including flyby-velocity and seasonal penalties, and $T$ is the total duration of the partial tour. The second used a weighted combination of total unbonused score and unbonused score rate,
\begin{equation}
    R_c = \lambda_w J + \lambda_r \frac{J}{T}
\end{equation}
which allowed high-value structures with longer durations to remain in the beam without allowing very slow branches to dominate. The third used the unbonused score rate with an explicit grand-tour bonus,
\begin{equation}
    R_g = \frac{J}{T} + \lambda_g I_{\mathrm{GT}}
\end{equation}
where $I_{\mathrm{GT}}$ indicates whether the partial tour had completed the major-planet portion of the grand tour requirement. These different objectives were useful because they produced a diverse set of solutions and emphasized different parts of the search space.

\begin{figure}[t]
    \centering
    \includegraphics[width=3.25in]{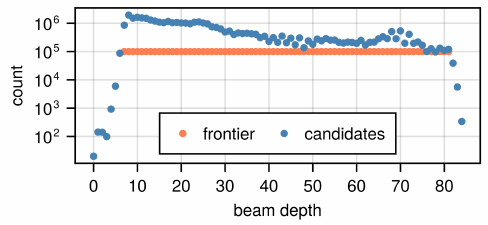}
    \caption{Progress statistics throughout beam search.}
    \label{fig:beam_expansion}
\end{figure}

\begin{figure}[t]
    \centering
    \includegraphics[width=\textwidth]{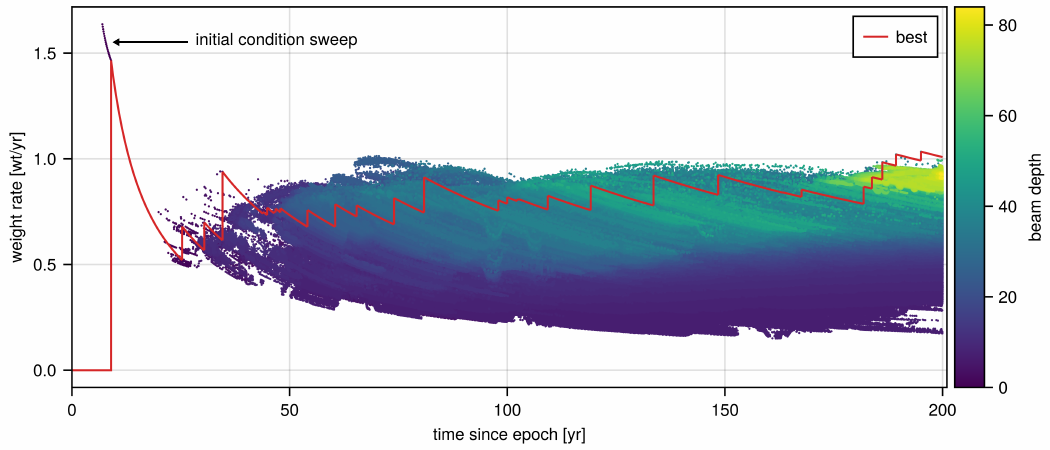}
    \caption{Distribution of candidate beam search tours throughout the beam search.}
    \label{fig:beam_weight_scatter}
\end{figure}

The search also retains an elite set containing the best complete or partial tours encountered over all depths. These elites were stored separately from the current frontier and are ranked independently on the unbonused score. This is important because a useful sequence may leave the frontier as the depth increases, even if it remains valuable for later refinement or analysis. The relationship between the candidates and the retained frontier is shown in Fig.~\ref{fig:beam_expansion}. The number of candidates is generally larger than the retained frontier size because each frontier node can produce multiple feasible child nodes. In the example shown, the frontier size is fixed at 100,000 nodes, while the candidate set can be more than an order of magnitude larger. Eventually, the search terminates when no remaining frontier nodes produce valid children.

\begin{figure}[t]
    \centering
    \includegraphics[width=\textwidth]{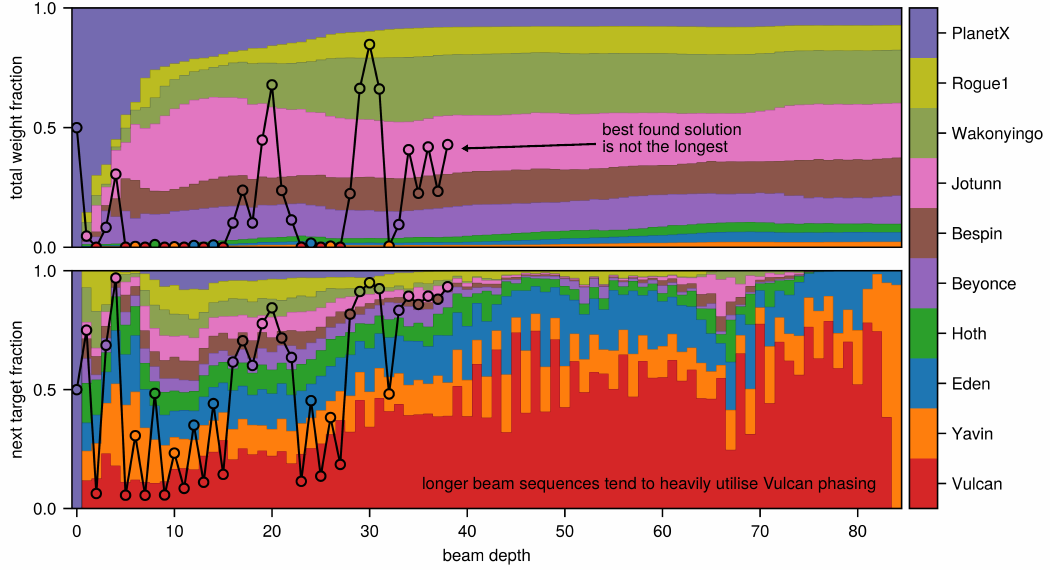}
    \caption{Planetary encounter statistics used to assess beam search gravity assist structures.}
    \label{fig:beam_frontier_analysis}
\end{figure}

The resulting distribution of candidates throughout an example beam search is shown in Fig.~\ref{fig:beam_weight_scatter}. In this case, the unbonused-score-rate objective $R_w$ is used. The starting nodes are visible as high score-rate points because of the large initial score associated with the PlanetX encounter. As the beam depth increases, the candidate tours generally become longer, and the pruning process gradually removes low-scoring branches, causing the search to focus on local expansions of more promising sequences. The best elite node is highlighted in red. Its path through the search demonstrates why retaining a large set of nodes is important: the best solution was not always the best-ranked node at intermediate depths, and therefore would not have been found by a purely greedy search.

Further information on the behavior of the beam search is shown in Fig.~\ref{fig:beam_frontier_analysis}. The maximum depth for this search was 84 moves. As the search progresses, the distribution of the weight of the nodes in the frontier is shown in the top panel. The bottom panel then shows the distribution of the possible next moves across the frontier. Most of the score in the best beam-search candidates comes from flybys of the outer planets throughout the entire search depth. It is important to note that this indicates that the beam search frontier contains many sequences that visit all of the outer planets. Therefore, the outputs of the planetary beam search could be candidate structures for satisfying the major-planet part of the grand tour requirement.

\subsection{Ballistic Tour Solutions}
\label{sec:ballistic_tour}

\begin{figure}[t]
    \centering
    \includegraphics[width=\textwidth]{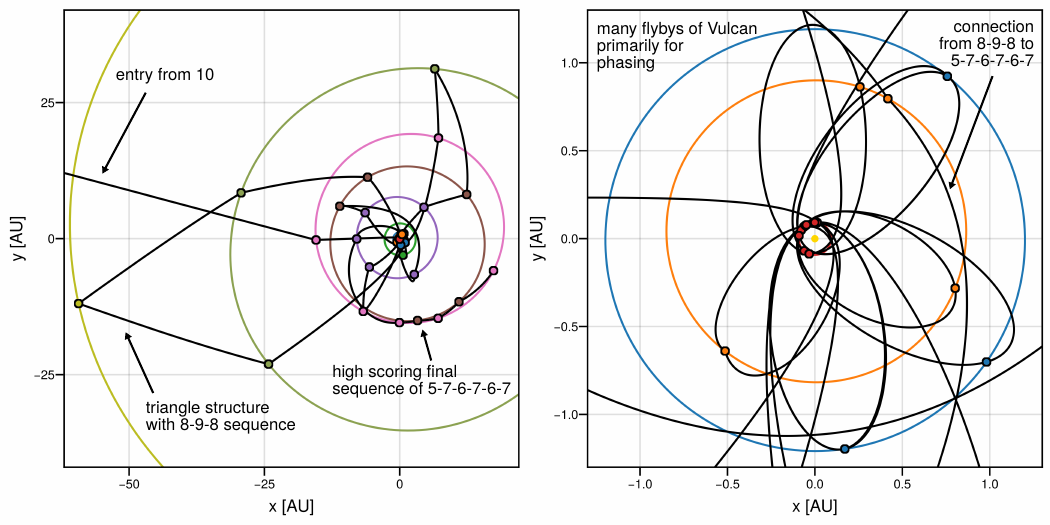}
    \caption{Representative ballistic tour identified by the beam search.}
    \label{fig:beam_trajectory}
\end{figure}

A wide range of beam searches for major-planet \gls{ga} tours were run with different frontier sizes and objectives. Frontier sizes from 50,000 to 2,500,000 were tested, producing a variety of high-quality solutions. These runs generally found trajectories with total unbonused scores between 200 and 220. The analysis resulting from these investigations identified several different high-scoring planetary tour structures. A representative ballistic tour that visits all major planets is shown in Fig.~\ref{fig:beam_trajectory}.

This tour gives a useful perspective on a potential grand-tour strategy. Since PlanetX is visited on the entry leg, the remaining major-planet requirement can be completed by a sequence that visits all other planets. The key geometry is an $8$-$9$-$8$ triangular structure, which scores highly and visits Rogue1, followed by a return toward the inner system and a fast, high-scoring $5$-$7$-$6$-$7$-$6$-$7$ chain. The inner planets mainly serve as phasing opportunities because their raw scientific weights are comparatively low. Together, these features form an approximately 100-year terminal sequence that both completes the outer-planet portion of the grand tour and provides a comparatively high score. This gives important insight into the periods of time in which other strategies, such as the construction of resonant tours, could be most effective.

\subsection{Finding Resonant Tours}
\label{sec:resonant_design}

The location of Vulcan close to Altaira, together with its large gravitational parameter, provides a strong opportunity to use repeated \glspl{ga} around the central star. This is particularly attractive when combined with the low periapsis of several comets and the increased effectiveness of the solar sail near Altaira. These features motivate a strategy in which Vulcan is used as a patch point between repeated comet flybys. Since comets have three times the raw scientific weight of asteroids, this could produce a high-scoring strategy even when the flyby-velocity penalty is large, because the time required to visit each comet can be small.

Near Altaira, there are two primary mechanisms that can change the trajectory: the solar sail and \glspl{ga} at Vulcan. The solar sail alone does not provide enough control authority to efficiently construct resonant tours on a competitive timescale. However, when a \gls{ga} at Vulcan places the spacecraft into a Keplerian orbit, that orbit is guaranteed to intersect Vulcan's orbit again at a later time. If the orbital period is resonant with the period of Vulcan, then a subsequent Vulcan \gls{ga} can be performed. These resonant return arcs can be selected so that they also intersect the trajectory of a comet or other object, enabling a link between consecutive Vulcan flybys.

This motivates a strategy that is parameterized in terms of the Vulcan orbital period, $P_V$. Let $R=P_K/P_V$ denote the ratio between the Keplerian period of the spacecraft orbit and the Vulcan orbital period. For a rational ratio $R=p/q$, written in lowest terms, the spacecraft and Vulcan return to the same relative phasing after $q$ spacecraft revolutions and $p$ Vulcan periods. Let $N_{\mathrm{ret}}(R)=p$ denote this number of Vulcan periods. The Vulcan-return duration, $T_R$, is then
\begin{equation}
    T_R = N_{\mathrm{ret}}(R)P_V
    \label{eq:resonant-return-time}
\end{equation}
This definition admits fractional resonant ratios. For example, $R=22.5=45/2$ corresponds to a return after two spacecraft revolutions and 45 Vulcan periods. Within this analysis, the candidate ratios are chosen such that $N_{\mathrm{ret}}(R)\leq150$, so the spacecraft returns to Vulcan within 150 Vulcan periods. The maximum direct resonant-transfer duration is therefore $T_R=150P_V=1500$ days.

The variables $\alpha_1$ and $\alpha_2$ are then introduced to place intermediate flybys within a fixed Vulcan-return interval. For a direct Vulcan-target-Vulcan option, the target flyby is parameterized by $\alpha_1\in(0,1)$ and occurs at $\tau=t_i+\alpha_1T_R$. Let $m=(b_{\mathrm{retro}},N_{\mathrm{rev}},b_{\mathrm{long}})$ denote the Lambert branch flags. The Keplerian period ratio produced by the outgoing Vulcan-target Lambert arc is
\begin{equation}
    R_K(\tau;t_i,j,m)
    =
    \frac{
        P_K(\boldsymbol{r}_i,\boldsymbol{v}_{\mathrm{out},j}(\tau;t_i,m))
    }{P_V}
\end{equation}
where $\boldsymbol{v}_{\mathrm{out},j}(\tau;t_i,m)$ is the Vulcan departure velocity and $P_K$ is the Keplerian period of the resulting orbit. A residual can then be constructed as the difference between the achieved and desired Keplerian period ratios,
\begin{equation}
    g_{\mathrm{dir},j}(\tau;t_i,R,m)
    =
    R_K(\tau;t_i,j,m)-R
    \label{eq:resonant-direct-mismatch}
\end{equation}

\begin{figure}[t]
    \centering
    \includegraphics[width=\textwidth]{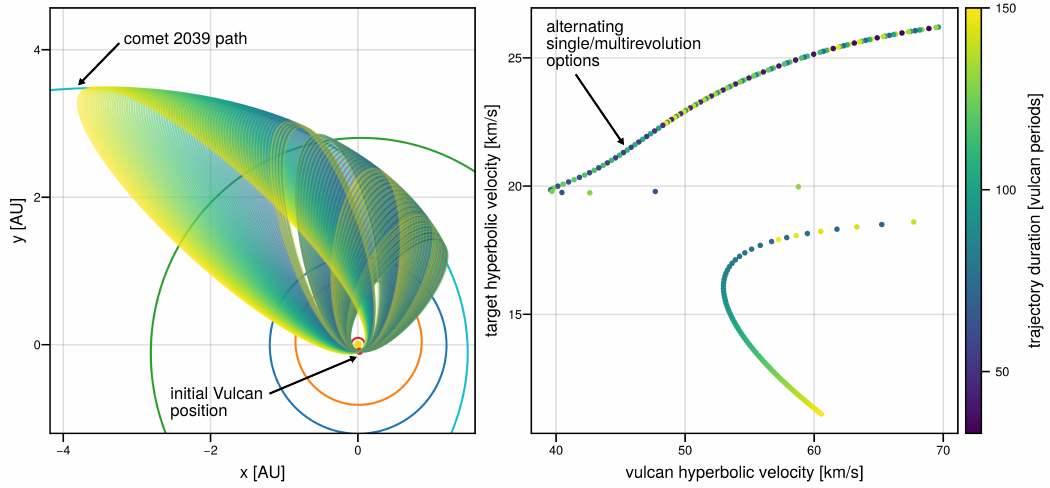}
    \caption{Simplified resonant Vulcan-return geometry used for comet-targeting sequences.}
    \label{fig:resonant_simple}
\end{figure}

Roots of this equation can then be found using a similar strategy to Section~\ref{sec:ballistic_ga}, where a one-dimensional scan is performed over the fractional flyby time $\alpha_1$ using a fixed number of brackets. Brackets of Eq.~\eqref{eq:resonant-direct-mismatch} that contain a sign change are then refined with either the \texttt{ITP} \citep{oliveiraEnhancementBisectionMethod2021} or \texttt{modAB} \citep{ganchovskiImprovementsModifiedAnderson2026} method. A total of 100 brackets are used for $\alpha_1\in(0.1,0.9)$. The conservative limits help to prevent cases where a flyby occurs too close to Vulcan, which makes it difficult to further refine using the \gls{scp} formulation introduced in Section~\ref{sec:scp}. Finally, candidate options are removed if any resulting Keplerian arc violates the general periapsis limit of 0.05 AU.

This concept is illustrated in Fig.~\ref{fig:resonant_simple} for comet 2039. The spacecraft begins at Vulcan at a fixed time, and a scan is conducted over all possible resonant period ratios. For each target object, a Lambert arc is constructed from Vulcan to the target. The target flyby time is adjusted until the resulting Keplerian orbit has the desired resonant period ratio. The corresponding return time is then fixed by Eq.~\eqref{eq:resonant-return-time}. Repeating this process produces a map of possible Vulcan departure velocities and cometary target-relative arrival velocities, demonstrating the many resonant flyby options that exist over a wide range of transfer durations and Vulcan flyby velocities.

This process can also be extended to include planetary \glspl{ga}, since cometary transfers may not always be available and some planetary sequences can also score well. For a single intermediate \gls{ga}, such as a Vulcan-Eden-Vulcan sequence, the return duration $T_R$ is still fixed by the scanned resonant ratio. However, the trajectory is split into two Keplerian arcs joined at the intermediate body. Instead of requiring the incoming and outgoing flyby-velocity vectors at the intermediate body to match exactly, the root solve enforces equality of their magnitudes. Because this signed mismatch can be positive or negative, the same bracketing methods can be used. A turning-angle check is then applied to determine whether the resulting \gls{ga} is feasible. A periapsis check is also required for both Keplerian arcs, since the \gls{ga} changes the orbital elements at the intermediate body.

For a single intermediate \gls{ga} at body $j_1$, the \gls{ga} time is again parameterized by $\alpha_1\in(0.1,0.9)$ and occurs at $\tau_1=t_i+\alpha_1T_R$. For a Lambert branch pair $(m_1,m_2)$, the incoming and outgoing flyby velocities are computed from the adjacent Lambert arcs using Eq.~\eqref{eq:ga-vinf-def}. The signed flyby-speed mismatch is the same quantity defined in Eq.~\eqref{eq:beam-vinf-mismatch}, and the single-\gls{ga} root condition is
\begin{equation}
    g_{j_1}(\alpha_1;t_i,T_R,m_1,m_2)=0,
    \label{eq:resonant-single-ga-mismatch}
\end{equation}
which satisfies the unpowered flyby-speed condition in Eq.~\eqref{eq:ga-speed-match}. The corresponding turn angle is then checked using Eqs.~\eqref{eq:ga-turning-angle} and \eqref{eq:ga-delta-feasible}.

\begin{figure}[t]
    \centering
    \includegraphics[width=\textwidth]{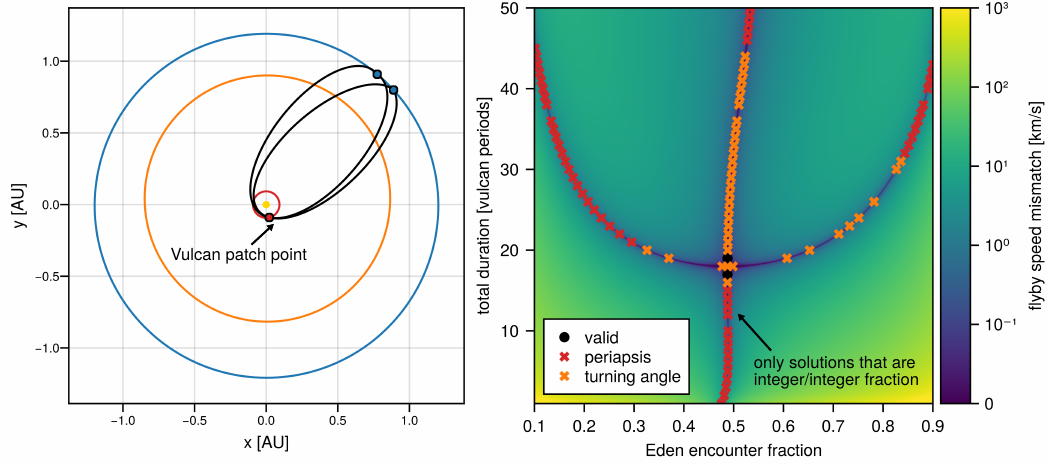}
    \caption{Single-GA resonant transfer timing map for a Vulcan-Eden-Vulcan sequence.}
    \label{fig:single_ga}
\end{figure}

An example of this process is shown in Fig.~\ref{fig:single_ga} for the construction of a Vulcan-Eden-Vulcan sequence. As the scanned resonant ratio changes, the return duration $T_R$ changes and the root locations shift across the fractional encounter time. Candidate solutions are obtained whenever a root exists for the selected ratio, and infeasible cases are removed using the periapsis and turning-angle constraints. In this example, two feasible solutions remain, representing possible planetary \gls{ga} sequences that could be included within the Vulcan-resonant search.

\begin{figure}[t]
    \centering
    \includegraphics[width=\textwidth]{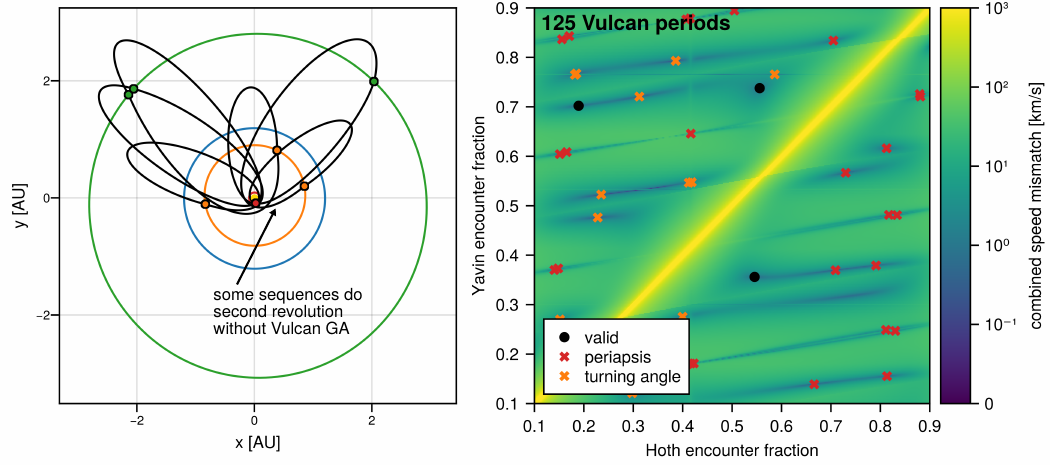}
    \caption{Double-GA resonant transfer search for Vulcan-return sequences.}
    \label{fig:double_ga}
\end{figure}

\begin{figure}[t]
    \centering
    \includegraphics[width=\textwidth]{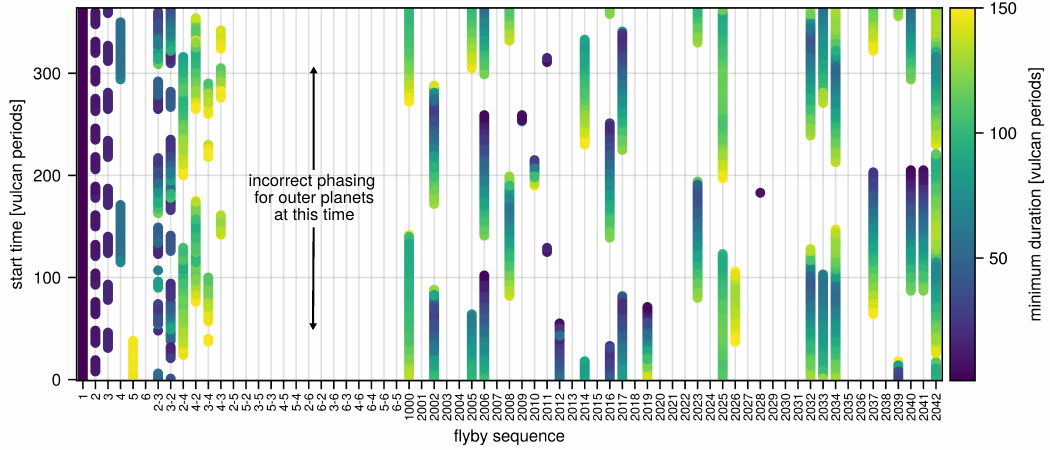}
    \caption{Resonant-transfer timing map for candidate Vulcan-return tours.}
    \label{fig:resonant_timeframe_map}
\end{figure}

\begin{algorithm}[!t]
\caption{Generation of resonant transfer options.}
\label{alg:resonant_option_generation}
\KwInput{Vulcan departure epoch $t_i$, ephemerides $\boldsymbol{e}$, candidate ratios $\mathcal{R}$, target-time brackets $\mathcal{B}_\tau$, fraction brackets $\mathcal{B}_\alpha$, Lambert branches $\mathcal{M}$, reference period $P_V$}
\KwOutput{Resonant options $\mathcal{O}_i$ for departure epoch $t_i$}
$\mathcal{O}_i\gets\emptyset$\;
\ForEach{ratio $R\in\mathcal{R}$}{
    $T_R\gets N_{\mathrm{ret}}(R)P_V$\;
    Store feasible self-resonant Vulcan option for ratio $R$ in $\mathcal{O}_i$\tcp*{waiting options}
    \ForEach{direct target $j\in\mathcal{J}_{\mathrm{direct}}$}{
        \ForEach{Lambert branch $m\in\mathcal{M}$}{
            $\mathcal{Z}\gets\mathrm{FindRoots}(g_{\mathrm{dir},j}(\tau;t_i,R,m),\mathcal{B}_\tau)$\tcp*{direct roots Eq.~\eqref{eq:resonant-direct-mismatch}}
            \ForEach{$\tau^\ast\in\mathcal{Z}$}{
                $o\gets\mathrm{DirectOption}(t_i,j,\tau^\ast,R,m)$\;
                Store $o$ in $\mathcal{O}_i$ if $\mathrm{Feasible}(o)$\;
            }
        }
    }
    \ForEach{assist body $j_1\in\mathcal{J}_{\mathrm{GA}}$}{
        \ForEach{branch pair $(m_1,m_2)\in\mathcal{M}^2$}{
            $\mathcal{Z}\gets\mathrm{FindRoots}(g_{j_1}(\alpha_1;t_i,T_R,m_1,m_2),\mathcal{B}_\alpha)$\tcp*{single-\gls{ga} roots Eq.~\eqref{eq:resonant-single-ga-mismatch}}
            \ForEach{$\alpha_1^\ast\in\mathcal{Z}$}{
                $o\gets\mathrm{AssistOption}(t_i,[j_1],[\alpha_1^\ast T_R],R,[m_1,m_2])$\;
                Store $o$ in $\mathcal{O}_i$ if $\mathrm{Feasible}(o)$\;
            }
        }
    }
    \ForEach{ordered assist pair $(j_1,j_2)$ from $\mathcal{J}_{\mathrm{GA}}$}{
        \ForEach{branch triple $(m_1,m_2,m_3)\in\mathcal{M}^3$}{
            $\mathcal{Z}\gets\mathrm{FindGridRoots}(\boldsymbol{g}_{j_1,j_2}(\alpha_1,\alpha_2;t_i,T_R,m_1,m_2,m_3),\mathcal{B}_\alpha^2)$\tcp*{double-\gls{ga} roots Eq.~\eqref{eq:resonant-double-ga-mismatch}}
            \ForEach{$(\alpha_1^\ast,\alpha_2^\ast)\in\mathcal{Z}$}{
                $o\gets\mathrm{AssistOption}(t_i,[j_1,j_2],[\alpha_1^\ast T_R,\alpha_2^\ast T_R],R,[m_1,m_2,m_3])$\;
                Store $o$ in $\mathcal{O}_i$ if $\mathrm{Feasible}(o)$\;
            }
        }
    }
}
\KwReturn $\mathcal{O}_i$\;
\end{algorithm}

In principle, this process could be extended in a similar manner to sequences containing any number of \glspl{ga}. However, within this analysis, the search was limited to at most two intermediate \glspl{ga}. For two intermediate \glspl{ga} at bodies $j_1$ and $j_2$, the encounter times are parameterized by $\alpha_1$ and $\alpha_2$, with $0.1<\alpha_1<\alpha_2<0.9$, so that $\tau_1=t_i+\alpha_1T_R$ and $\tau_2=t_i+\alpha_2T_R$. For a Lambert branch triple $(m_1,m_2,m_3)$, the root solve is applied to the vector of the two signed flyby-speed mismatches,
\begin{equation}
    \boldsymbol{g}_{j_1,j_2}(\alpha_1,\alpha_2;t_i,T_R,m_1,m_2,m_3)
    =
    \begin{bmatrix}
        g_{j_1}(\alpha_1,\alpha_2;t_i,T_R,m_1,m_2)\\
        g_{j_2}(\alpha_1,\alpha_2;t_i,T_R,m_2,m_3)
    \end{bmatrix}
    =
    \boldsymbol{0}.
    \label{eq:resonant-double-ga-mismatch}
\end{equation}
For double-\gls{ga} sequences, this makes the search space two-dimensional. Therefore, direct bracketing methods, as in the direct and single-\gls{ga} cases, are much less effective. Instead, a coarse grid of possible encounter timings is generated, and local refinement using Newton's method is initialized from promising grid points. The Newton solve is formed from the two flyby-speed mismatches, one for each intermediate \gls{ga}. Finally, the turning-angle constraints are checked at both intermediate bodies, and the periapsis constraint is checked on all three Keplerian arcs. An example of the results of this process is shown in Fig.~\ref{fig:double_ga}, where the combined flyby-velocity mismatch is plotted for a fixed resonant return duration. Compared to the direct and single-\gls{ga} cases, the resulting objective landscape is now much more complex, and many solutions begin to fail the periapsis and turning-angle constraints.

The direct, single-\gls{ga}, and double-\gls{ga} searches are then combined to create a catalog of possible resonant transfers from Vulcan at a given epoch. Purely self-resonant transfers are also included so that there is always an available option at all possible Vulcan \gls{ga} times. Each possible flyby sequence is considered for Vulcan-resonant periods up to 150 Vulcan periods. The resulting option-generation algorithm for the direct, single-\gls{ga}, and double-\gls{ga} cases is summarized in Algorithm~\ref{alg:resonant_option_generation}.

Repeating this scan over the possible Vulcan \gls{ga} epochs produces the transfer map shown in Fig.~\ref{fig:resonant_timeframe_map}. The scan includes all comets, the dwarf planet Yandi, and planetary sequences involving planets up to Planet 6, including double-\gls{ga} combinations between them. The evolution of the transfer opportunities over time is clearly visible. Some planets are useful periodically, for example Planet 5 near the start, while a wide range of comet opportunities appear periodically with different minimum transfer durations.

\subsection{Beam Search for Resonant Tours}
\label{sec:resonance}

The resonant option catalog describes locally feasible Vulcan-return options for individual departure epochs. Each option contains a resonant ratio $R_o$, duration $T_R(o)$, intermediate target sequence, target-relative flyby velocities, the required outgoing Vulcan flyby velocity $\boldsymbol{v}_{\infty,V}^{+,o}$, and the incoming flyby velocity $\boldsymbol{v}_{\infty,V}^{-,o}$ produced at the next return. A complete resonant tour must then connect these options through consecutive Vulcan \glspl{ga} while preserving the incoming and outgoing flyby geometry at each return.

This structure naturally defines a graph-search problem, where each node is a Vulcan \gls{ga} and each edge is a possible resonant transfer option. However, there are several difficulties with applying a direct graph search to this map. First, it is not always possible to obtain the Vulcan flyby velocity required to depart onto a selected option. Second, even if the flyby velocity magnitudes are compatible, the required turning angle at Vulcan may not be achievable. This introduces a history dependence, since whether an option can be taken depends on the incoming flyby velocity produced by the previous option. For this reason, direct graph search methods such as a \gls{mip} are difficult to apply. Instead, a beam search was again used, with the resonant option catalog defining the possible moves at each Vulcan \gls{ga}.

In this beam search, additional pruning checks are applied when connecting a chain node to a candidate option. At a node $n$ within the beam, the stored velocity $\boldsymbol{v}_{\infty,V}^{-,n}$ is the incoming Vulcan flyby velocity produced by the previous option. A candidate option requires the outgoing flyby velocity $\boldsymbol{v}_{\infty,V}^{+,o}$. Instead of requiring the magnitudes of these velocities to match, which would be very difficult to ensure, this difference is relaxed to only require similar magnitudes. The idea is that the post beam-search refinement using \gls{scp} will resolve these \glspl{ga} by using the solar sail and by changing the time of the \gls{ga}.

Thresholds of approximately 20 km/s for this difference were found to work well and produced solutions that successfully refined using the \gls{scp}, as shown in Section~\ref{sec:resonance_results}. Finally, the required turning angle between the two flyby velocities at Vulcan is checked using the minimum and maximum feasible periapsis radii, as in Eq.~\eqref{eq:ga-delta-feasible}. The maximum and minimum of the two flyby velocities are used in this check to produce a conservative approximation.

After these pruning steps, the beam search ranks the remaining child nodes using an objective and retains the best candidates for further expansion. In this case, the best-weight-rate objective in Eq.~\eqref{eq:beam-max-weight-rate} is used. Each retained node then carries its own incoming Vulcan flyby velocity, allowing the search to account for the history dependence of the resonant sequence. This process is summarized in Algorithm~\ref{alg:resonant_beam_search}.

\begin{algorithm}[ht]
\caption{Beam search over resonant transfer options.}
\label{alg:resonant_beam_search}
\KwInput{Initial frontier $\mathcal{F}_0$, epoch-indexed resonant option catalog $\mathcal{O}$, frontier size $K$, elite count $K_E$, node ranking objective $\Phi_R$, Vulcan flyby speed tolerance $\epsilon_v$}
\KwOutput{Elite resonant chains $\mathcal{E}$}
$\mathcal{F}\gets\mathcal{F}_0$, $\mathcal{E}\gets\mathcal{F}_0$\tcp*{frontier nodes are Vulcan encounters}
\While{$\mathcal{F}\neq\emptyset$}{
    $\mathcal{C}\gets\emptyset$\;
    \ForEach{chain node $n=(t_n,T^n,W^n,H^n,\boldsymbol{v}_{\infty,V}^{-,n},Q^n)\in\mathcal{F}$}{
        $\mathcal{L}_n\gets\emptyset$\tcp*{local successors from node $n$}
        \ForEach{option $o\in\mathcal{O}_{t_n}$}{
            Skip if option $o$ is blocked or targets a blocked body\;
            Reject if $t_n+T_R(o)$ lies outside the epoch grid\;
            Reject if not $\mathrm{VulcanFeasible}(\boldsymbol{v}_{\infty,V}^{-,n},\boldsymbol{v}_{\infty,V}^{+,o},\epsilon_v)$\tcp*{Vulcan speed/turning-angle}
            $\Delta W\gets\sum_{j\in\mathrm{targets}(o)} w_j F(\|\boldsymbol{v}_{\infty,j}^{-,o}\|)S(\hat{\boldsymbol{r}}_j,H^n)$\tcp*{score gained by flybys}
            $H'\gets\mathrm{Append}(H^n,\mathrm{targets}(o))$\;
            $n_{\mathrm{child}}\gets(t_n+T_R(o),T^n+N_{\mathrm{ret}}(R_o),W^n+\Delta W,H',\boldsymbol{v}_{\infty,V}^{-,o})$
            $Q'\gets\Phi_R(n_{\mathrm{child}})$\;
            Append child state $n_{\mathrm{child}}$ with score $Q'$ to $\mathcal{L}_n$\;
        }
        Keep the best local successors of $n$ by ranking objective and append them to $\mathcal{C}$\;
    }
    \If{$\mathcal{C}=\emptyset$}{
        \KwBreak\;
    }
    $\mathcal{E}\gets\operatorname{TopK}(\mathcal{E}\cup\mathcal{C},K_E)$ by total resonant score\;
    $\mathcal{F}\gets\operatorname{TopK}(\mathcal{C},K)$ by $\Phi_R$\;
}
\KwReturn $\mathcal{E}$\;
\end{algorithm}

\subsection{Resonant Tour Solutions}
\label{sec:resonant_tour}

\begin{figure}[t]
    \centering
    \includegraphics[width=\textwidth]{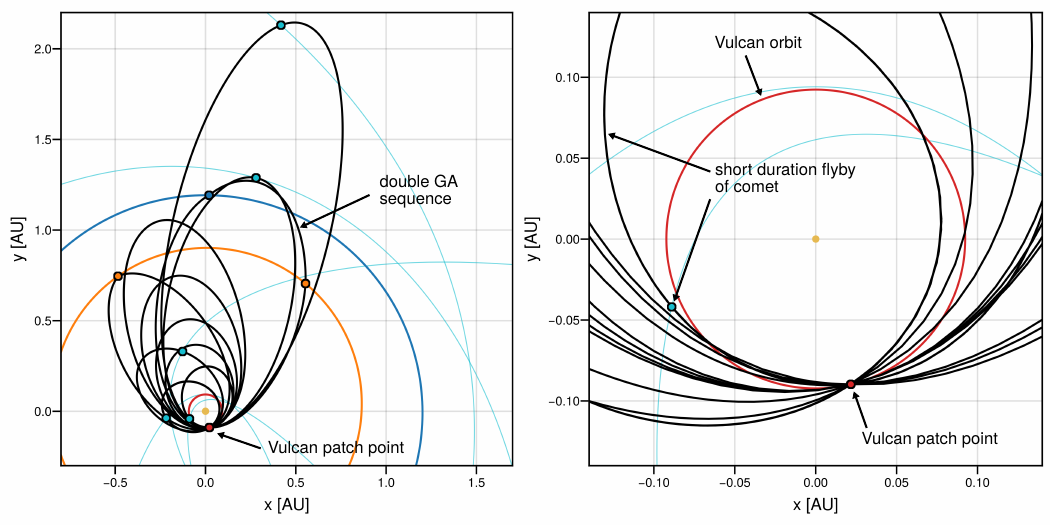}
    \caption{Example Vulcan-resonant tour solution generated with beam search.}
    \label{fig:resonant_tour_beam}
\end{figure}

An example resonant chain produced by the beam search is shown in Fig.~\ref{fig:resonant_tour_beam}, with the corresponding option sequence listed in Table~\ref{tab:resonant_ballistic_sequence}. This is a ballistic resonant sequence rather than a fully refined solar sail trajectory. The incoming velocity at the first Vulcan encounter is $51.40$ km/s. In the table, $v_{\infty}^{+}$ is the outgoing flyby speed required by the selected option, $v_{\infty,\mathrm{next}}^{-}$ is the incoming flyby speed produced at the next return, $\Delta v_\infty$ is the speed mismatch used in the Vulcan-feasibility pruning check, and $\delta$ is the corresponding turn angle.

\begin{table}[t]
\centering
\caption{Details on the example Vulcan-resonant tour solution generated with beam search.}
\begin{tabular}{rlrrrrrrr}
\toprule
Depth & Target(s) & TOF [$P_V$] & TOF [yr] & $v_{\infty}^{+}$ [km/s] & $v_{\infty,\mathrm{next}}^{-}$ [km/s] & $\Delta v_\infty$ [km/s] & $\delta$ [deg] & Score \\
\midrule
1 & 2039 & 44 & 1.20 & 52.94 & 52.94 & 1.54 & 67.9 & 1.08 \\
2 & 2 & 11 & 0.30 & 40.17 & 37.52 & 12.77 & 84.8 & 0.44 \\
3 & 2012 & 5 & 0.14 & 48.12 & 48.12 & 10.60 & 40.5 & 0.60 \\
4 & 1 & 10 & 0.27 & 33.78 & 33.78 & 14.34 & 65.9 & 0.00 \\
5 & 2019 & 5 & 0.14 & 48.13 & 48.13 & 14.35 & 75.3 & 0.60 \\
6 & 2017 & 21 & 0.57 & 44.43 & 44.43 & 3.70 & 70.1 & 1.00 \\
7 & 1 & 6 & 0.16 & 30.45 & 30.45 & 13.97 & 37.3 & 0.00 \\
8 & 2006 & 7 & 0.19 & 44.42 & 44.42 & 13.96 & 77.6 & 0.60 \\
9 & 1 & 16 & 0.44 & 35.94 & 35.94 & 8.48 & 77.5 & 0.00 \\
10 & 2,3 & 22 & 0.60 & 49.06 & 41.13 & 13.12 & 48.1 & 1.21 \\
11 & 1 & 4 & 0.11 & 26.79 & 26.79 & 14.34 & 29.2 & 0.00 \\
\midrule
$\Sigma$ & 12 total & 151 & 4.13 & -- & -- & -- & -- & 5.55 \\
\bottomrule
\end{tabular}
\label{tab:resonant_ballistic_sequence}
\end{table}

The sequence visits 12 targets over 151 Vulcan periods, corresponding to 4.13 years, and accumulates an unbonused score of 5.55. This indicates a potential score rate of 1.34 per year, which is very comparable to the best portions of the ballistic planetary beam-search results in Section~\ref{sec:ballistic_tour}. However, the Vulcan flyby-speed mismatches are not zero, so this solution should be considered to be a high-quality initial structure for the solar sail refinement rather than a directly valid ballistic tour.

\section{Solar Sail Tour Design} \label{sec:solar_sail_design}

The ballistic search results from Section~\ref{sec:ballistic_design} provide useful insight into potential structures for a grand tour solution. This includes the identification of high-scoring ballistic planetary tours, as well as the identification of resonant Vulcan-return sequences. However, these sequences have mismatched flyby velocities at Vulcan and cannot be used directly. Therefore, a fast and effective methodology for using the solar sail is required.

This section describes the trajectory optimization framework developed for the \gls{gtoc13} problem. A general \gls{scp} formulation is introduced for the range of trade studies and refinement problems considered in this work. This formulation uses a lossless convexification of the solar sail control constraints \citep{oguriLosslessControlConvexFormulation2024}, together with the trajectory constraints and objectives needed for refinement. The formulation is then applied to several design tasks: the refinement of ballistic planetary tours, the assessment of initial entry options, the refinement of terminal planetary sequences, the assessment of asteroid-belt tours, and the transformation of resonant Vulcan-return sequences into feasible trajectories.

The parameterization for the solar sail within the \gls{scp} uses the standard cone and clock angles. This reduces the three-component normal vector to two independent control variables, although it requires the creation of a rotating reference frame attached to the spacecraft. The cone angle $\alpha$ is the angle between the inward radial direction and the solar sail normal, and the clock angle $\beta$ defines the rotation of the solar sail normal about the radial direction in the local transverse plane, which is defined based on the ecliptic direction.

In this case, the sail normal vector $\hat{\boldsymbol{u}}_n$ is expressed in an Altaira-pointing rotating frame $F_{\mathrm{AN}}=(\hat{\boldsymbol{r}},\hat{\boldsymbol{k}}_t,\hat{\boldsymbol{k}}_z)$ centered on the spacecraft. The construction of the unit vectors of this frame is defined through the sequence of operations
\begin{align}
    \hat{\boldsymbol{r}} &= -\hat{\boldsymbol{u}}_r \\
    \hat{\boldsymbol{k}}_z &=
    \frac{
        \hat{\boldsymbol{z}}^*-(\hat{\boldsymbol{z}}^*\cdot\hat{\boldsymbol{u}}_r)\hat{\boldsymbol{u}}_r
    }{
        \left\|\hat{\boldsymbol{z}}^*-(\hat{\boldsymbol{z}}^*\cdot\hat{\boldsymbol{u}}_r)\hat{\boldsymbol{u}}_r\right\|_2
    } \\
    \hat{\boldsymbol{k}}_t &= \hat{\boldsymbol{r}}\times\hat{\boldsymbol{k}}_z
    \label{eq:sail-frame}
\end{align}
where $\hat{\boldsymbol{u}}_r$ is the unit vector from the spacecraft to Altaira, and $\hat{\boldsymbol{z}}^*$ is the normal vector to the ecliptic plane. Then, the sail unit normal $\hat{\boldsymbol{u}}_n$ is expressed in the rotating frame $F_{\mathrm{AN}}$ as a function of the cone angle $\alpha$ and clock angle $\beta$, which are derived from a sequence of rotations about the local axes of the frame:
\begin{equation}
    \hat{\boldsymbol{u}}_n
    =
    \begin{bmatrix}
        -\cos\alpha \\
        \sin\alpha\sin\beta \\
        \sin\alpha\cos\beta
    \end{bmatrix}_{F_{\mathrm{AN}}}
    \label{eq:sail-normal}
\end{equation}

\begin{figure}[t]
    \centering
    \includegraphics[width=3.25in]{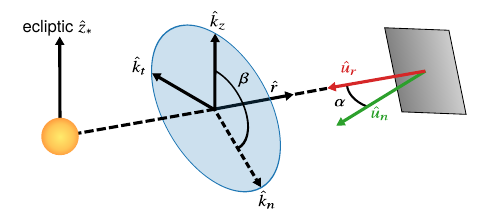}
    \caption{Definition of the cone angle $\alpha$ and clock angle $\beta$ in the local Altaira-pointing frame $F_{\mathrm{AN}}$ for the ideal flat-plate solar sail.}
    \label{fig:srp_frame}
\end{figure}

These angles and the geometric construction of the unit vectors are illustrated in Fig.~\ref{fig:srp_frame}. The cone angle is restricted to $\alpha\in[0^\circ,90^\circ]$, so the solar sail always faces Altaira, while the clock angle is unrestricted. At a cone angle of $90^\circ$ there is no acceleration from the solar sail and the dynamics match the ballistic model.

\subsection{Sequential Convex Programming} \label{sec:scp}
A critical component of the \gls{gtoc13} solution process is an optimization algorithm for solar sailing spacecraft. In this work, the trajectory refinement is performed using a convex optimization approach based on \gls{scp}. The resulting method computes the solar sail control profile, encounter times, and \gls{ga} geometry required to transform a prescribed target sequence into an optimized solar sailing trajectory. The core idea behind \gls{scp} is to replace the full nonlinear trajectory optimization problem with a sequence of convex approximate subproblems. Each subproblem is constructed about the current reference trajectory, and the resulting solution is then used to update the reference for the next iteration. Under suitable conditions, this process converges to a local optimum of the original nonlinear problem \citep{malyutaConvexOptimizationTrajectory2022}. Although many subproblems may need to be solved, each one is convex and can be solved reliably and efficiently \citep{boydConvexOptimization2004}. This makes \gls{scp} a practical approach for large trajectory refinement problems that would be difficult to solve directly using large nonlinear programming formulations.

Following the general \gls{scp} approach \citep{malyutaConvexOptimizationTrajectory2022}, the first step is to construct a convex approximation of the dynamics and define any additional constraints for a single trajectory leg. In this work, a leg is defined as the segment between two consecutive trajectory events, such as the initial approach, a planetary \gls{ga}, or a flyby of a comet or asteroid. The trajectory within each leg is discretized into nodes and segments, and an optional zero-order-hold solar sail control is assigned to each segment. The dynamics are then enforced between adjacent nodes, while path and event constraints can be applied at the relevant nodes of the leg.

This formulation is then extended to multi-leg trajectory problems. By introducing constraints that connect legs together, many legs can be solved simultaneously as a single \gls{scp} problem. These connection constraints include massless-body flybys, patched-conic \glspl{ga}, and time-continuity constraints. By framing the problem in this way, the same formulation can be used for a variety of related problems, including short local refinements, phase connection problems, and the final large-scale refinement of complete tour sequences.

\subsubsection{Leg Constraints}
\label{sec:leg_constraints}

Each trajectory leg is formulated using a direct transcription of the spacecraft dynamics \citep{conwaySpacecraftTrajectoryOptimization2010}. The state at node $j$ is defined as $\boldsymbol{x}_j=(\boldsymbol{r}_j,\boldsymbol{v}_j)$, and the zero-order-hold solar sail control over the following segment is denoted by $\boldsymbol{u}_j=(u_{j,x},u_{j,y},u_{j,z})$. The node times $t_j$ may also be included as decision variables, allowing the optimizer to adjust flyby epochs and, when required, the underlying time mesh \citep{kumagaiAdaptiveMeshSequentialConvex2024}. The dynamics in Eq.~\eqref{eq:dynamics-original} are then reformulated in terms of this control vector as
\begin{equation}
    \dot{\boldsymbol{x}}
    =
    \begin{bmatrix}
        \dot{\boldsymbol{r}}(t) \\
        \dot{\boldsymbol{v}}(t)
    \end{bmatrix}
    =
    \begin{bmatrix}
        \boldsymbol{v} \\
        -\dfrac{\mu}{r^3}\boldsymbol{r}
        + \dfrac{\beta}{r^2}\hat{\boldsymbol{a}}_{\mathrm{sail}}(\boldsymbol{u})
    \end{bmatrix}
    \label{eq:dynamics}
\end{equation}
where $\beta = 2(C A / m)\, r_{\mathrm{ref}}^2$ is a lumped solar-radiation-pressure parameter for the sail. The normalized acceleration direction $\hat{\boldsymbol{a}}_{\mathrm{sail}}$ is determined from the control vector $\boldsymbol{u}_j$ in the Altaira-pointing frame $F_{\mathrm{AN}}=(\hat{\boldsymbol{r}},\hat{\boldsymbol{k}}_t,\hat{\boldsymbol{k}}_z)$ as
\begin{equation}
    \hat{\boldsymbol{a}}_{\mathrm{sail},j}
    =
    u_{j,x}\hat{\boldsymbol{r}} - u_{j,z}\hat{\boldsymbol{k}}_t - u_{j,y}\hat{\boldsymbol{k}}_z
    \label{eq:sail-accel}
\end{equation}

\textbf{(linearized dynamics)}
The original nonlinear dynamics are linearized about a reference trajectory $\{\bar{\boldsymbol{x}}_j,\bar{\boldsymbol{u}}_j,\bar{t}_j\}$ to obtain a convex defect constraint for every segment:
\begin{equation}
    \boldsymbol{x}_{j+1}
    = \bar{\boldsymbol{x}}_{j+1}
    + \boldsymbol{A}_j(\boldsymbol{x}_j-\bar{\boldsymbol{x}}_j)
    + \boldsymbol{B}_j(\boldsymbol{u}_j-\bar{\boldsymbol{u}}_j)
    + \boldsymbol{C}_j(t_j-\bar{t}_j)
    + \boldsymbol{D}_j(t_{j+1}-\bar{t}_{j+1})
    + \boldsymbol{\sigma}_{j}
    \label{eq:constraint-dynamics}
\end{equation}
where $\boldsymbol{\sigma}_{j}$ is a virtual control used to preserve feasibility of the convex subproblem, $\boldsymbol{A}_j$ is the state transition matrix, $\boldsymbol{B}_j$ is the control sensitivity matrix, and $\boldsymbol{C}_j$ and $\boldsymbol{D}_j$ are the sensitivities with respect to the initial and terminal times of the segment. For legs with fixed node times, the $\boldsymbol{C}_j$ and $\boldsymbol{D}_j$ terms are omitted. For compact notation, the reference data for segment $j$ is defined as
\begin{equation}
    \mathcal{Y}_j =
    \left(
        \bar{\boldsymbol{x}}_j,\bar{\boldsymbol{u}}_j,\bar{t}_j,\bar{t}_{j+1}
    \right)
    \label{eq:reference-segment}
\end{equation}
The sensitivity matrices are then computed as
\begin{align}
    \boldsymbol{A}_j &= \left. \left[\frac{\partial}{\partial \boldsymbol{x}} \int_{t_j}^{t_{j+1}}\dot{\boldsymbol{x}}\,\text{d}t \right]\right|_{\mathcal{Y}_j}\\
    \boldsymbol{B}_j &= \left. \left[\frac{\partial}{\partial \boldsymbol{u}} \int_{t_j}^{t_{j+1}} \dot{\boldsymbol{x}}\,\text{d}t \right]\right|_{\mathcal{Y}_j}\\
    \boldsymbol{C}_j &= \left. \left[\frac{\partial}{\partial t_j} \int_{t_j}^{t_{j+1}} \dot{\boldsymbol{x}}\,\text{d}t \right]\right|_{\mathcal{Y}_j}\\
    \boldsymbol{D}_j &= \left. \left[\frac{\partial}{\partial t_{j+1}} \int_{t_j}^{t_{j+1}} \dot{\boldsymbol{x}}\,\text{d}t \right]\right|_{\mathcal{Y}_j}
\end{align}

Rather than deriving analytic expressions for these sensitivities, they are computed through the use of \gls{ad} directly on the numerical integration solver of each segment. In this work, the \texttt{Tsit5} integrator \citep{tsitourasRungeKuttaPairs2011}, as implemented in \texttt{DifferentialEquations.jl} \citep{rackauckasDifferentialEquationsJlPerformant2017}, is used for the segment propagation with relative and absolute tolerances of $10^{-12}$. Forward-mode \gls{ad} is performed using \texttt{ForwardDiff.jl} \citep{revelsForwardModeAutomaticDifferentiation2016}.

\begin{figure}[t]
    \centering
    \includegraphics[width=3.25in]{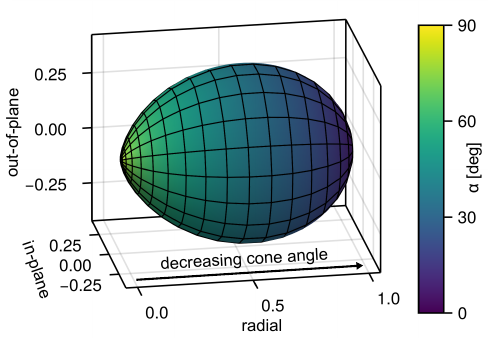}
    \caption{Ideal flat-plate solar sail acceleration envelope as a function of sail orientation.}
    \label{fig:acceleration_map}
\end{figure}

\textbf{(lossless solar sail)}
Direct use of the cone and clock angles as optimization variables generally leads to poor linearization performance because the feasible set of solar sail acceleration vectors is nonlinear and nonconvex. This set is visualized in Fig.~\ref{fig:acceleration_map} for the ideal flat-plate solar sail model used in the \gls{gtoc13} problem. For this reason, the \gls{scp} formulation uses a lossless convexification of the solar sail control constraints \citep{oguriLosslessControlConvexFormulation2024}, adapted here to the ideal flat-plate model. The approach consists of two components: a relaxed convex representation of the admissible acceleration region, and a small objective penalty on $u_x$ that drives the relaxation to bind at optimality. After optimization, the resulting acceleration components can then be mapped back to the corresponding cone and clock angles. The relaxed feasible set is enforced by

\begin{figure}[t]
    \centering
    \includegraphics[width=3.25in]{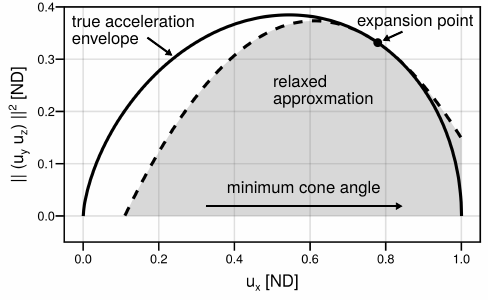}
    \caption{Lossless convexified solar sail control region used in the SCP subproblems.}
    \label{fig:lossless_region}
\end{figure}

\begin{align}
    \cos^3\alpha_\text{max} &\leq u_{x,j} \leq \cos^3\alpha_\text{min} \label{eq:constraint-sail-radial} \\
    u_{yz,j} &\geq \left\|(u_{y,j},u_{z,j})\right\|_2 \quad (\text{SOC})\label{eq:constraint-sail-soc} \\
    u_{yz,j} &\leq h(\bar{u}_{x,j}) + h'(\bar{u}_{x,j})(u_{x,j}-\bar{u}_{x,j})
    + \tfrac{1}{2}h''(\bar{u}_{x,j})(u_{x,j}-\bar{u}_{x,j})^2 \label{eq:constraint-sail-taylor}
\end{align}
where
\begin{align}
    \tau(u_x) &= u_x^{1/3} \\
    h(u_x) &= \tau^2\sqrt{1-\tau^2}
    \label{eq:sail-h-definition}
\end{align}
First, the radial acceleration component $u_x$ is bounded by the maximum and minimum cone angles, $\alpha_\text{max}$ and $\alpha_\text{min}$. In this work, a $0.5^\circ$ buffer is added to make $\alpha_\text{min}=0.5^\circ$ and $\alpha_\text{max}=89.5^\circ$. This prevents numerical instabilities arising from large gradients in the derivative approximation near $u_x=0$ and $u_x=1$ in Fig.~\ref{fig:lossless_region}, and does not substantially change the solution. The transverse acceleration components $u_y$ and $u_z$ are then bounded by a \gls{soc} constraint. Finally, the upper bound on the transverse acceleration is enforced through a second \gls{soc} constraint, which uses a quadratic, but strictly convex, approximation of Eq.~\eqref{eq:sail-h-definition}. The derivatives of $h$ are computed using \gls{ad} for simplicity and the approximation is recomputed around the reference at each \gls{scp} iteration. A small objective term, $-\lambda_c\sum_j u_{x,j}$ with $\lambda_c=10^{-3}$, encourages the relaxed solution to lie on the boundary of the true solar sail acceleration set by favoring smaller cone angles. An illustration of the resulting relaxed control region is shown in Fig.~\ref{fig:lossless_region}.

\textbf{(leg boundary conditions)}
Each leg may include different initial and terminal boundary constraints depending on its role in the tour. First, the \gls{gtoc13} entry condition, which constrains the spacecraft to the prescribed approach geometry, can be expressed as:
\begin{equation}
    \begin{bmatrix}
        x_1-r_{\text{approach}} \\
        v_{y,1} \\
        v_{z,1}
    \end{bmatrix}
    =
    \boldsymbol{0}
    \label{eq:constraint-initial}
\end{equation}
where the superscripts indicate the component of the vector and $r_{\text{approach}}=-200~\text{AU}$. This constrains the initial condition of the spacecraft to the approach plane, as well as the requisite incoming velocity directions to zero, while allowing the remaining components of the initial state to be optimized within the \gls{scp}. Depending on the problem, the velocity component $v_{x,1}$ can be fixed to a specified approach speed or left as an optimization variable.

For legs that begin or end at a target with a specified ephemeris, the boundary constraints match the corresponding body state. When the event time is optimized, the ephemeris state is linearized about the reference event time. For example, the terminal state constraint becomes
\begin{align}
    \boldsymbol{r}_{\text{end}}
    &=
    \bar{\boldsymbol{r}}_{\text{end}}
    + \bar{\boldsymbol{v}}_{\text{end}}(t_{\text{end}}-\bar{t}_{\text{end}})
    + \boldsymbol{\sigma}_{\text{end}}^{(r)}
    \label{eq:constraint-final-position} \\
    \boldsymbol{v}_{\text{end}}+\Delta\boldsymbol{v}_{\text{end}}
    &=
    \bar{\boldsymbol{v}}_{\text{end}}
    + \bar{\boldsymbol{a}}_{\text{end}}(t_{\text{end}}-\bar{t}_{\text{end}})
    + \boldsymbol{\sigma}_{\text{end}}^{(v)}
    \label{eq:constraint-final-velocity}
\end{align}
where $\boldsymbol{\sigma}_{\text{end}}^{(r)}$ and $\boldsymbol{\sigma}_{\text{end}}^{(v)}$ are boundary virtual controls for the position and velocity matching constraints. Analogous constraints can be used for an initial state constraint. These constraints also introduce the velocity terms $\Delta\boldsymbol{v}_0$ and $\Delta\boldsymbol{v}_\text{end}$, which are used to support flybys. The magnitudes of these terms are bounded with \gls{soc} constraints:
\begin{align}
    \|\Delta\boldsymbol{v}_0\|_2 &\leq \Delta v_{0} \quad (\text{SOC})\label{eq:constraint-flyby-velocity-bounds1}\\
    \|\Delta\boldsymbol{v}_\text{end}\|_2 &\leq \Delta v_{\text{end}} \quad (\text{SOC})
    \label{eq:constraint-flyby-velocity-bounds2}
\end{align}
where the non-bold symbols $\Delta v_0$ and $\Delta v_{\text{end}}$ are scalar upper bounds on the corresponding flyby velocities.

\textbf{(leg timing constraints)}
The node times can be fixed, bounded, or optimized adaptively. For example, they can be used to limit the final node time to the maximum permitted by the \gls{gtoc13} problem. In general, these constraints take the form:
\begin{align}
    t_1 &\geq t_{\min}\label{eq:constraint-time-bounds1}\\
    t_{\text{end}} &\leq t_{\max}
    \label{eq:constraint-time-bounds2}
\end{align}
where $t_{\min}$ and $t_{\max}$ are the minimum and maximum times, respectively. These can also be used in a similar manner to limit the maximum duration of a leg or sequence of legs. Another important constraint to consider is to limit the maximum expansion or contraction that a segment may make. This helps to prevent segments from being forced to very small or very large timespans, which can lead to numerical issues. These are expressed relative to the reference mesh:
\begin{equation}
    \frac{1}{\eta_c}
    \leq
    \frac{t_{j+1}-t_j}{\bar{t}_{j+1}-\bar{t}_j}
    \leq
    \eta_e
    \label{eq:constraint-time-sequential}
\end{equation}
where $\eta_c$ and $\eta_e$ represent the minimum and maximum contraction and expansion factors, respectively.

\textbf{(minimum radius constraint)}
A minimum-radius constraint is included in some refinement phases to prevent close approaches to Altaira. Since the constraint $\|\boldsymbol{r}_j\|_2 \geq r_{\min}$ is nonconvex, it is linearized about the reference position $\bar{\boldsymbol{r}}_j$:
\begin{align}
    r_{\min} &\leq \|\bar{\boldsymbol{r}}_j\|_2
    +
    \frac{\bar{\boldsymbol{r}}_j^\top}{\|\bar{\boldsymbol{r}}_j\|_2}
    \left(\boldsymbol{r}_j-\bar{\boldsymbol{r}}_j\right)
    \label{eq:constraint-min-radius}
\end{align}

\textbf{(trust regions)}
Fixed-size trust regions are imposed on the state, and on the node times when required,
\begin{align}
    -\varepsilon_x \leq \boldsymbol{x}_j-\bar{\boldsymbol{x}}_j \leq \varepsilon_x \label{eq:constraint-trust-state}\\
    -\varepsilon_t \leq t_j-\bar{t}_j \leq \varepsilon_t \label{eq:constraint-trust-time}
\end{align}
These bounds limit the validity region of the linearized dynamics, ephemeris constraints, and event constraints within each \gls{scp} subproblem. The values of $\varepsilon_x$ and $\varepsilon_t$ are generally tuned manually depending on the problem requirements. Adaptive trust-region algorithms can work well in some situations \citep{maoSuccessiveConvexificationNonConvex2017,oguriSuccessiveConvexificationFeasibility2023}, but tend to be very sensitive to tuning parameters and can often prematurely converge.

\textbf{(virtual controls)}
Finally, the absolute value of the virtual controls is computed through auxiliary variables,
\begin{align}
    |\boldsymbol{\sigma}_j| &\geq \boldsymbol{\sigma}_j\label{eq:constraint-virtual1}\\
    |\boldsymbol{\sigma}_j| &\geq -\boldsymbol{\sigma}_j
    \label{eq:constraint-virtual2}
\end{align}

\subsubsection{Leg Linking Constraints}
\label{sec:multi_leg_constraints}

A complete tour can be represented by a sequence of $L$ trajectory legs connected through link constraints. The leg sequence and approximate timing are supplied, while the \gls{scp} formulation optimizes the node states, solar sail controls, flyby velocities, and flyby timing. Let $l\in\{1,\dots,L-1\}$ index the link from leg $l$ to leg $l+1$, with the leg index written as the final subscript when node or boundary indices are also present. The linking constraints define how each adjacent pair of legs should connect, whether through a massless-body flyby or a \gls{ga}.

\textbf{(state link)}
For massless bodies, including Yandi, asteroids, and comets, the flyby cannot change the spacecraft velocity. The massless-body link therefore enforces position and velocity continuity, while also requiring the terminal and initial flyby velocity variables to be equal:
\begin{align}
    \boldsymbol{r}_{\text{end},l} &= \boldsymbol{r}_{1,l+1} + \boldsymbol{\sigma}_{1,l+1}^{(r)} \label{eq:constraint-link-position} \\
    \boldsymbol{v}_{\text{end},l} + \Delta\boldsymbol{v}_{\text{end},l}
        &= \boldsymbol{v}_{1,l+1} + \Delta\boldsymbol{v}_{0,l+1}
        + \boldsymbol{\sigma}_{1,l+1}^{(v)} \label{eq:constraint-link-velocity} \\
    \Delta\boldsymbol{v}_{\text{end},l} &= \Delta\boldsymbol{v}_{0,l+1} \label{eq:constraint-link-dv}
\end{align}
If the terminal and flyby velocity variables are forced to zero, this would cause rendezvous. The boundary virtual controls $\boldsymbol{\sigma}_{1,l+1}^{(r)}$ and $\boldsymbol{\sigma}_{1,l+1}^{(v)}$ are therefore used to preserve feasibility of the convex subproblem and are penalized with the other dynamical virtual controls in Eq.~\eqref{eq:objective-penalty}.

\textbf{(time link)}
Time continuity between adjacent legs is enforced by
\begin{equation}
    t_{\text{end},l}=t_{1,l+1}
    \label{eq:constraint-link-time}
\end{equation}

\textbf{(gravity assist link)}
For planetary encounters, the massless-body link can be replaced by a \gls{ga} link. The position and time remain continuous, while the incoming and outgoing velocities must satisfy the patched-conic \gls{ga} model introduced in Section~\ref{sec:problem_statement}. For the \gls{ga} link after leg $l$, let $\boldsymbol{\xi}_l=[\xi_{1,l},\xi_{2,l},\xi_{3,l}]^\top$ denote virtual controls for the \gls{ga}. The flyby-speed equality in Eq.~\eqref{eq:ga-speed-match} is imposed in the convex subproblem by projecting the terminal and initial flyby correction variables onto the reference incoming $\hat{\boldsymbol{u}}_{\text{in}}$ and outgoing $\hat{\boldsymbol{u}}_{\text{out}}$ directions:
\begin{equation}
    \hat{\boldsymbol{u}}_{\text{in}}^\top \Delta\boldsymbol{v}_{\text{end},l}
    =
    \hat{\boldsymbol{u}}_{\text{out}}^\top \Delta\boldsymbol{v}_{0,l+1}
    + \xi_{1,l}
    \label{eq:constraint-ga-vinf}
\end{equation}
The scalar $\xi_{1,l}$ relaxes the linearized flyby-speed equality but is penalized in the objective function. The remaining components of $\boldsymbol{\xi}_l$, introduced below, play the same role for the upper and lower turning-angle constraints.

The \gls{ga} turning angle must also remain within the feasible range defined in Eq.~\eqref{eq:ga-delta-feasible}. Because the incoming and outgoing projected flyby speeds may differ in the convex subproblem, the lower turning-angle bound is evaluated using the smaller speed, while the upper turning-angle bound is evaluated using the larger speed:
\begin{align}
    \delta_{\min}(v_{\text{in}},v_{\text{out}})
    &=
    2\arcsin\!\left(\frac{1}{1+r_{p,\max}\min(v_{\text{in}},v_{\text{out}})^2/\mu_b}\right)\\
    \delta_{\max}(v_{\text{in}},v_{\text{out}})
    &=
    2\arcsin\!\left(\frac{1}{1+r_{p,\min}\max(v_{\text{in}},v_{\text{out}})^2/\mu_b}\right)
    \label{eq:constraint-ga-delta-bounds}
\end{align}
where $\mu_b$ is the gravitational parameter of the flyby planet. The projected incoming and outgoing speeds used in the convex subproblem are
\begin{align}
    v_{\text{in,lin}} =
    \hat{\boldsymbol{u}}_{\text{in}}^\top \Delta\boldsymbol{v}_{\text{end},l}\\
    v_{\text{out,lin}} =
    \hat{\boldsymbol{u}}_{\text{out}}^\top \Delta\boldsymbol{v}_{0,l+1}
    \label{eq:constraint-ga-speed-proj}
\end{align}

The actual turning angle between the incoming and outgoing flyby velocity vectors is
\begin{equation}
    \delta(\Delta\boldsymbol{v}_{\text{end},l},\Delta\boldsymbol{v}_{0,l+1})
    =
    \arccos\!\left(
    \frac{
    \Delta\boldsymbol{v}_{\text{end},l}^{\top}\Delta\boldsymbol{v}_{0,l+1}}
    {\|\Delta\boldsymbol{v}_{\text{end},l}\|_2\|\Delta\boldsymbol{v}_{0,l+1}\|_2}
    \right)
    \label{eq:constraint-ga-angle}
\end{equation}
The lower bound, upper bound, and turning angle are then all linearized about the reference encounter:
\begin{align}
    \delta_{\min,\text{lin}}
    &=
    \bar{\delta}_{\min}
    + \nabla\delta_{\min}^{\top}
    \begin{bmatrix}
        v_{\text{in,lin}}-\bar{v}_{\text{in}}\\
        v_{\text{out,lin}}-\bar{v}_{\text{out}}
    \end{bmatrix}\\
    \delta_{\max,\text{lin}}
    &=
    \bar{\delta}_{\max}
    + \nabla\delta_{\max}^{\top}
    \begin{bmatrix}
        v_{\text{in,lin}}-\bar{v}_{\text{in}}\\
        v_{\text{out,lin}}-\bar{v}_{\text{out}}
    \end{bmatrix}\\
    \delta_{\text{lin}}
    &=
    \bar{\delta}
    + \nabla\delta^{\top}
    \left(
    \begin{bmatrix}
        \Delta\boldsymbol{v}_{\text{end},l}\\
        \Delta\boldsymbol{v}_{0,l+1}
    \end{bmatrix}
    -
    \begin{bmatrix}
        \Delta\bar{\boldsymbol{v}}_{\text{end},l}\\
        \Delta\bar{\boldsymbol{v}}_{0,l+1}
    \end{bmatrix}
    \right)
    \label{eq:constraint-ga-linearized}
\end{align}
where the gradients are computed using \gls{ad} at the reference encounter. As a result, the sequential \gls{ga} turning-angle constraints are then
\begin{align}
    \delta_{\text{lin}}+\xi_{2,l}
    &\leq
    \delta_{\max,\text{lin}} \label{eq:constraint-ga-turning1}\\
    \delta_{\min,\text{lin}}
    &\leq
    \delta_{\text{lin}}+\xi_{3,l}
    \label{eq:constraint-ga-turning2}
\end{align}
The \gls{ga} virtual controls $\boldsymbol{\xi}_l$ are penalized more strongly than the dynamical virtual controls, so that violations of the patched-conic flyby model are resolved first.

\subsubsection{Objective Set}
\label{sec:objectives}

All objectives include penalty terms from the leg constraints and link constraints. The baseline penalty contribution is
\begin{equation}
    J_{\text{pen}} =
    \lambda \sum_{l=1}^{L}\sum_{j=1}^{N_l}\|\boldsymbol{\sigma}_{j,l}\|_1
    + \lambda_{\mathrm{GA}}\sum_{l=1}^{L-1}\|\boldsymbol{\xi}_{l}\|_1
    - \lambda_c\sum_{l=1}^{L}\sum_{j=1}^{N_l-1}u_{x,l,j}
    \label{eq:objective-penalty}
\end{equation}
where $\boldsymbol{\sigma}_{j,l}$ is the dynamical virtual control at node $j$ of leg $l$, and $\boldsymbol{\xi}_l$ collects the virtual controls used by the \gls{ga} link after leg $l$. The absolute values in the $L_1$ norms are obtained through the introduction of auxiliary variables. The weights $\lambda$ and $\lambda_{\mathrm{GA}}$ penalize violation of the linearized dynamics and \gls{ga} link constraints, respectively. Within this analysis, values of $\lambda=10^3$ and $\lambda_{\mathrm{GA}}=10^5$ tended to work well and provide the desired convergence behavior. The coefficient $\lambda_c$ is the small lossless solar sail binding weight, typically set to $10^{-3}$.

\textbf{(maximum weight)}
The primary competition objective maximizes the penalized science score defined in Eq.~\eqref{eq:score}. Both the flyby-velocity penalty and the seasonal penalty are nonlinear functions of the optimization variables, since they depend on the terminal flyby correction variable and the flyby location. These score terms are therefore linearized about the reference trajectory. Let $\mathcal{L}_s \subseteq \{1,\dots,L\}$ be the set of scoring legs, and let $W_l$ denote the full score contribution of leg $l$, including the target weight, flyby-velocity penalty, and seasonal penalty. The objective contribution is then
\begin{equation}
    J_w
    =
    -\sum_{l \in \mathcal{L}_s}
    \left[
        W_l(\bar{\boldsymbol{z}}_l)
        +
        \nabla W_l(\bar{\boldsymbol{z}}_l)^\top
        \left(\boldsymbol{z}_l-\bar{\boldsymbol{z}}_l\right)
    \right]
    \label{eq:objective-maxweight}
\end{equation}
where $\boldsymbol{z}_l$ collects the variables required to evaluate the score contribution for leg $l$, typically including $\Delta\boldsymbol{v}_{\text{end},l}$ and the flyby position of all previous visits to the target object. The gradients are again computed using \gls{ad} due to their complex formulations.

\textbf{(minimum duration)}
The minimum duration objective reduces the total time-of-flight of the entire trajectory:
\begin{equation}
    J_d = t_{L,\text{end}} - t_{1,1}
    \label{eq:objective-duration}
\end{equation}

\textbf{(minimum initial and terminal flyby velocity)}
The initial and terminal flyby velocity objectives minimize the velocity correction required to connect a leg to its neighboring phase or target body. The corresponding objective terms are
\begin{align}
    J_{v_{\infty,0}} = \Delta v_0 \\
    J_{v_{\infty,\text{end}}} = \Delta v_{\text{end}}
    \label{eq:objective-flyby-velocity}
\end{align}

\textbf{(minimum specific energy)}
The minimum-specific-energy objective minimizes the final orbital specific energy,
\begin{equation}
    \varepsilon_f = \frac{1}{2}\boldsymbol{v}_{L,\text{end}}^\top\boldsymbol{v}_{L,\text{end}}
    - \frac{\mu}{\|\boldsymbol{r}_{L,\text{end}}\|_2}
\end{equation}
Since this expression is nonlinear, it is linearized about the terminal reference state, again using \gls{ad} to compute the gradient:
\begin{equation}
    J_\varepsilon =
    \varepsilon(\bar{\boldsymbol{x}}_{L,\text{end}})
    +
    \nabla\varepsilon(\bar{\boldsymbol{x}}_{L,\text{end}})^\top
    \left(\boldsymbol{x}_{L,\text{end}}-\bar{\boldsymbol{x}}_{L,\text{end}}\right)
    \label{eq:objective-energy}
\end{equation}

\subsubsection{Complete SCP Formulation}

The complete convex subproblem solved at each \gls{scp} iteration can be written compactly as an optimization over the node states, sail controls, event times, flyby correction variables, and virtual controls:
\begin{mini*}
    {}{J_{\text{obj}} + J_{\text{pen}}}{}{}
    \addConstraint{\eqref{eq:constraint-dynamics}}{}{\quad\text{(linearized dynamics)}}
    \addConstraint{\eqref{eq:constraint-sail-radial},\,\eqref{eq:constraint-sail-soc},\,\eqref{eq:constraint-sail-taylor}}{}{\quad\text{(sail-control constraints)}}
    \addConstraint{\eqref{eq:constraint-initial},\,\eqref{eq:constraint-final-position},\,\eqref{eq:constraint-final-velocity},\,\eqref{eq:constraint-flyby-velocity-bounds1},\,\eqref{eq:constraint-flyby-velocity-bounds2}}{}{\quad\text{(leg boundary conditions)}}
    \addConstraint{\eqref{eq:constraint-time-bounds1},\,\eqref{eq:constraint-time-bounds2},\,\eqref{eq:constraint-time-sequential}}{}{\quad\text{(leg timing constraints)}}
    \addConstraint{\eqref{eq:constraint-min-radius}}{}{\quad\text{(minimum-radius constraint)}}
    \addConstraint{\eqref{eq:constraint-trust-state},\,\eqref{eq:constraint-trust-time}}{}{\quad\text{(trust regions)}}
    \addConstraint{\eqref{eq:constraint-virtual1},\,\eqref{eq:constraint-virtual2}}{}{\quad\text{(virtual-control absolute values)}}
    \addConstraint{\eqref{eq:constraint-link-position},\,\eqref{eq:constraint-link-velocity},\,\eqref{eq:constraint-link-dv},\,\eqref{eq:constraint-link-time}}{}{\quad\text{(massless-body and time links)}}
    \addConstraint{\eqref{eq:constraint-ga-vinf},\,\eqref{eq:constraint-ga-turning1},\,\eqref{eq:constraint-ga-turning2}}{}{\quad\text{(gravity-assist links)}}
\end{mini*}
where $J_{\text{obj}}$ is one of the objectives defined in Section~\ref{sec:objectives} and $J_{\text{pen}}$ is the penalty contribution in Eq.~\eqref{eq:objective-penalty}. Constraints are added or removed depending on the specific refinement problem.

The \gls{scp} refinement repeatedly solves this convex subproblem, propagates the resulting states and solar sail controls through the nonlinear dynamics, and rebuilds all sequential constraints about the updated reference trajectory. In the implementation, \texttt{JuMP.jl} \citep{lubinJuMPRecentImprovements2023} is used to construct the \gls{socp} subproblems, and \texttt{Clarabel} \citep{Clarabel_2024} is used as the solver. The state and time trust-region sizes are initially set to large values, for example $\varepsilon_x=1.0$, before being manually tightened across a sequence of solves once the solution starts to exhibit oscillations. A final high-accuracy solve is used to produce solutions valid for submission. The problem size scales with the number of legs $L$ and the number of segments per leg, with typical multi-leg tours producing convex subproblems on the order of $10^3$--$10^5$ variables.

\subsection{Refinement of Ballistic Tours}

A natural first use of the \gls{scp} framework is to directly refine the ballistic planetary tours identified in Section~\ref{sec:ballistic_design}. This explores whether the flyby times of a ballistic tour can be improved, and what benefits the solar sail can provide, without changing the encounter sequence itself. The Lambert arcs from the ballistic search are used as the initial reference trajectory for the \gls{scp}, with the corresponding encounter times used as the initial time guesses. The objective is to maximize the total penalized scientific score of the trajectory in Eq.~\eqref{eq:objective-maxweight}.

\begin{table}[b]
\centering
\caption{Refinement options for a ballistic tour solution.}
\begin{tabular}{lrrr}
\toprule
Refinement option & $v_{x,0}$ [km/s] & TOF [yr] & Score \\
\midrule
baseline & 34.951 & 195.030 & 201.693 \\
flyby time & 34.951 & 193.096 & 204.823 \\
solar sail & 34.951 & 195.030 & 203.030 \\
flyby time + solar sail & 34.951 & 193.991 & 208.819 \\
\bottomrule
\end{tabular}
\label{tab:ballistic_refinement}
\end{table}

The results of this analysis are shown in Table~\ref{tab:ballistic_refinement}. The baseline ballistic trajectory has a score of 201.693 and a total time of flight of 195.030 years. Allowing the flyby times to vary within the \gls{scp} refinement increases the score to 204.823, with the primary effect being a reduction in the final arrival time, likely to improve the optimality of the final sequence of flybys. In contrast, if the flyby times are fixed and only the solar sail is enabled, the score also improves but by a smaller amount. As would be expected, the largest improvement is obtained when both the flyby timing and the solar sail control are optimized, increasing the score to 208.819. This corresponds to an approximate $4\%$ improvement over the baseline, with most of the gain coming from improved phasing of the terminal sequence.

\begin{figure}[t]
    \centering
    \includegraphics[width=\textwidth]{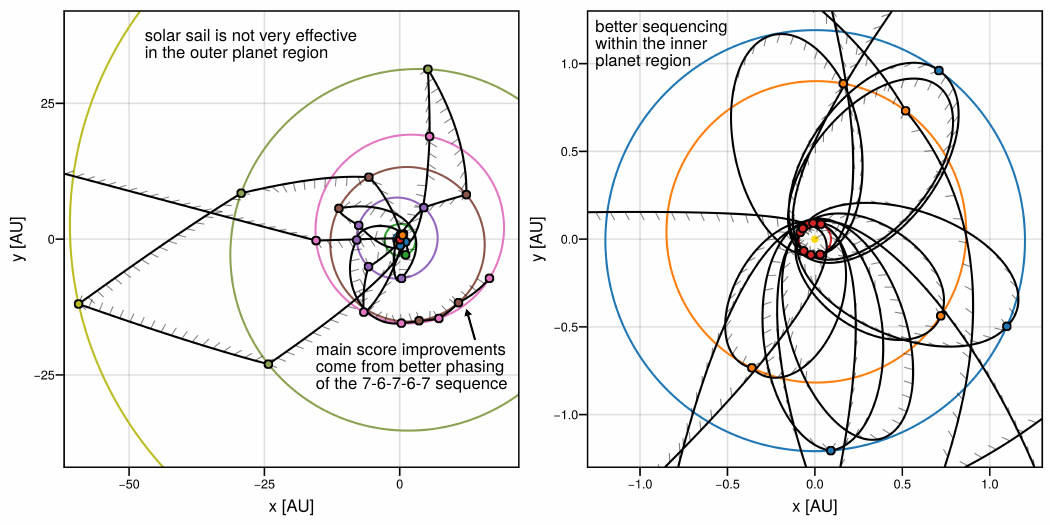}
    \caption{Direct SCP refinement of the beam search solution.}
    \label{fig:beam_planet_analysis}
\end{figure}

The refined trajectory is shown in Fig.~\ref{fig:beam_planet_analysis}. The optimized solar sail normal vector is illustrated throughout the trajectory. This refinement includes solar sail control on every segment, from the start of the trajectory to the end. For this refinement, a solar sail control discretization of 0.1 years was used, resulting in a total of 2010 nodes and 1971 segments across 39 legs. The resulting optimization problem has a total of 46,770 variables and 32,660 constraints. This optimization completed in less than one minute on a single thread of an AMD Ryzen 7 5800X3D processor, taking approximately 60 iterations for the \gls{scp} to converge. It is important to note that the runtime depends strongly on the choice of discretization included within the \gls{scp} problem, although the solution is not particularly sensitive to this choice. The selected discretization is sufficient to accurately capture the solar sail control and the resulting trajectory, while also keeping the problem size manageable.

\subsection{Initial Entry Analysis}
\label{sec:entry}

In the beam search, the initial entry was assumed to target PlanetX. This was a useful and ultimately correct decision, but the cost of this assumption needed to be quantified. In particular, a very different entry geometry, such as a direct close pass of Altaira designed to reduce the incoming energy as much as possible, could potentially have produced a better global solution.

The first part of this analysis estimates the maximum entry speed into the Altaira system that can still lead to capture. For this purpose, an \gls{scp} problem is constructed to minimize the terminal specific energy. The initial guess is generated from ballistic entry trajectories that satisfy the \gls{gtoc13} initial condition and reach the one-time minimum periapsis limit of 0.01 AU for a range of fixed entry velocities. An inverse-square node spacing was used along each trajectory, placing more nodes near the close approach where the solar sail acceleration and energy change are largest. This provided sufficient resolution for the refinement using only 400 nodes. Furthermore, to avoid solutions that have more than one revolution, the terminal time was set to occur only a few days after periapsis.

\begin{figure}[t]
    \centering
    \includegraphics[width=\textwidth]{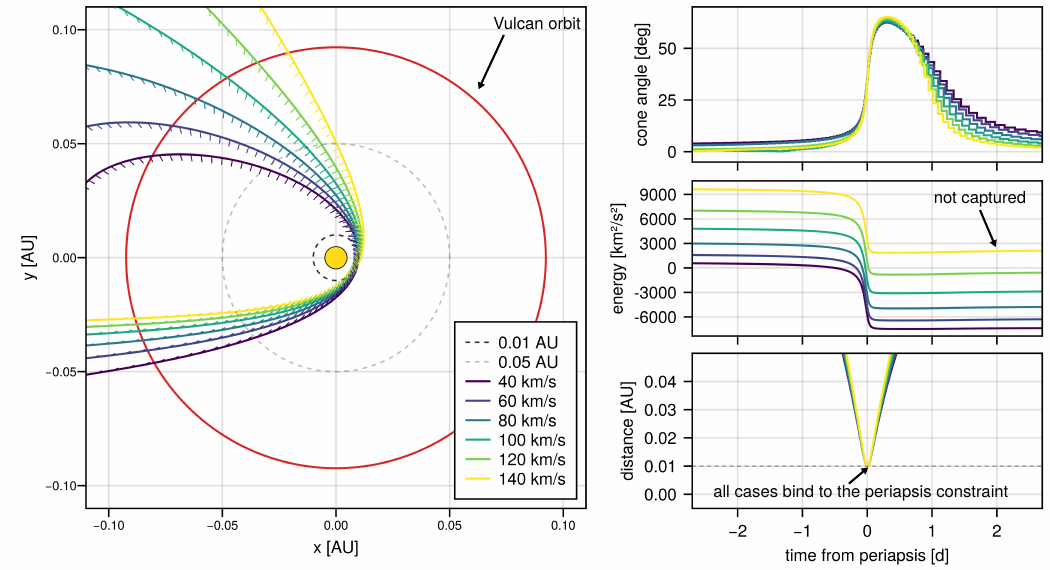}
    \caption{Solar sail refinement of selected entry options into the Altaira system.}
    \label{fig:entry_solar_sail}
\end{figure}

The fixed entry velocity results are shown in Fig.~\ref{fig:entry_solar_sail}, with time normalized about the periapsis epoch. For entry velocities below approximately 120 km/s, the solar sail can reduce the terminal specific energy below zero during the close pass of Altaira. These trajectories require approximately 8 years to reach periapsis. For lower entry velocities, the optimized trajectory is rapidly captured into a close orbit about Altaira, with an apoapsis approaching the orbit of Vulcan. For all cases, the cone-angle profiles are qualitatively similar, with the primary energy reduction occurring near periapsis. All cases bind to the minimum-radius constraint at periapsis.

This analysis indicates that direct capture is possible, with the limiting entry velocity lying between 120 and 140 km/s. To find the absolute maximum, the initial velocity was allowed to vary while enforcing a zero terminal specific energy constraint. This gave a maximum direct-capture entry velocity of 122.606 km/s, requiring 7.715 years to reach periapsis. If an optimally phased Vulcan flyby is included immediately after the close solar pass, an additional energy reduction is possible through the \gls{ga}. Under this assumption, the maximum entry velocity increases to 151.419 km/s, with a minimum duration of 6.251 years.

\begin{figure}[t]
    \centering
    \includegraphics[width=\textwidth]{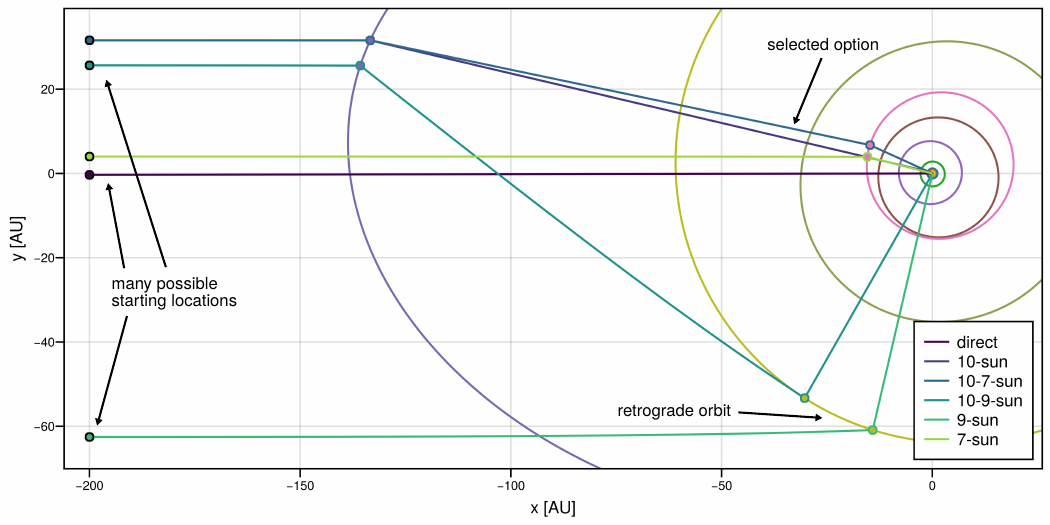}
    \caption{Assessment of ballistic entry options into the Altaira system.}
    \label{fig:entry_options}
\end{figure}

\begin{table}[t]
\centering
\caption{Possible entry sequence options for Altaira capture.}
\begin{tabular}{lrrrr}
\toprule
Entry sequence & $T_{\min}$ [yr] & $v_{x,0}$ [km/s] & Score & Score rate [yr$^{-1}$] \\
\midrule
sun & 7.715 & 122.606 & 0.000 & 0.000 \\
sun-1 & 6.251 & 151.419 & 0.000 & 0.000 \\
7-sun & 24.556 & 38.044 & 3.629 & 0.148 \\
9-sun & 65.502 & 16.581 & 13.966 & 0.213 \\
10-sun & 23.031 & 41.321 & 11.957 & 0.519 \\
10-7-sun & 22.748 & 41.825 & 15.368 & 0.676 \\
10-9-sun & 88.721 & 11.670 & 45.160 & 0.509 \\
\bottomrule
\end{tabular}
\label{tab:entry_weights}
\end{table}

These direct-capture options can then be compared with entry sequences that first use planetary flybys before the close pass of Altaira. These options are visualized in Fig.~\ref{fig:entry_options}, with the corresponding data summarized in Table~\ref{tab:entry_weights}. The selected entry option was the 10-7-sun sequence. This option provides a high score rate, enables the grand-tour bonus through the flyby of PlanetX, and improves on the 10-sun option because the additional Planet 7 \gls{ga} helps to turn the trajectory toward Altaira for the close solar pass.

The direct entry options were therefore less useful than initially expected. The grand-tour bonus is enabled by the flyby of PlanetX and is worth, in terms of available mission duration, approximately $(200-200/1.2)\approx 33$ yr. The direct entry option saves at most approximately 16 yr relative to the selected 10-7-sun entry, but scores zero and would not allow for the grand tour. To be competitive, the time saved by direct entry would need to compensate for both the 15.368 points scored by the selected entry sequence and the loss of the grand-tour bonus. Avoiding the grand-tour requirement could simplify the terminal sequence, particularly by removing the 8-9-8 triangle, but this would be attractive only if the resonant Vulcan phase in Section~\ref{sec:resonance} could achieve a substantially higher score rate and be extended over a longer duration.

\subsection{Refinement of Terminal Sequences}

\begin{figure}[b!]
    \centering
    \includegraphics[width=\textwidth]{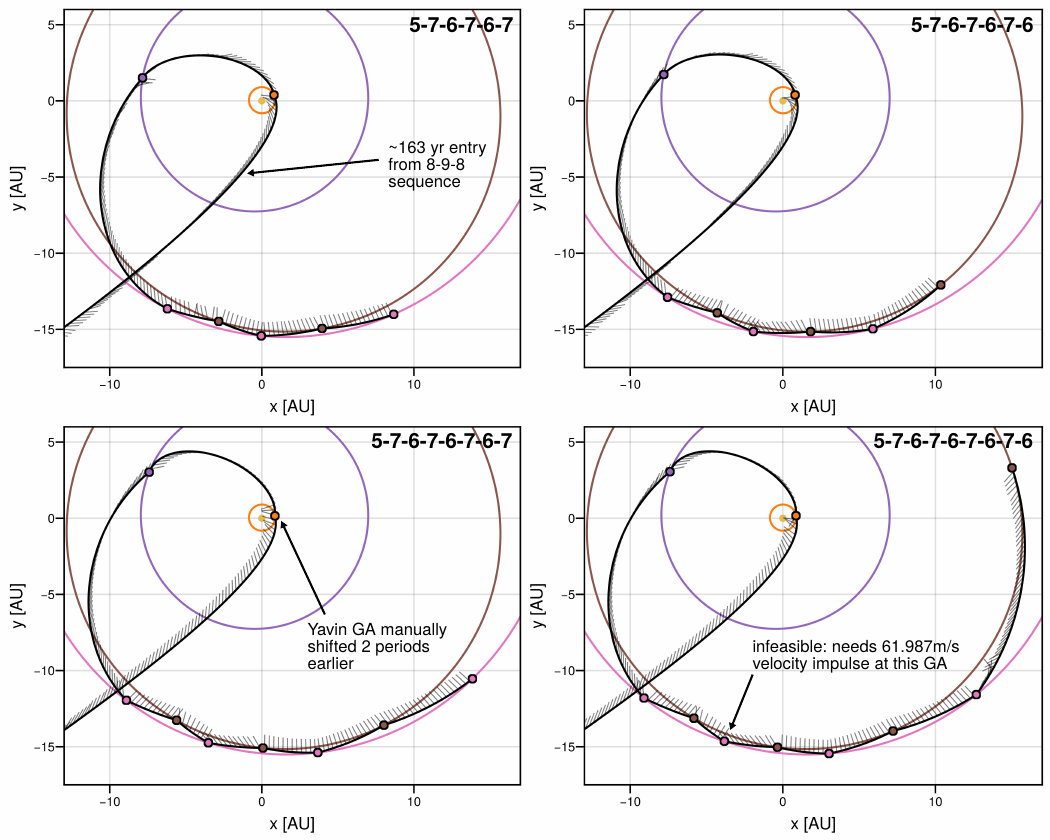}
    \caption{Refinement process of the terminal sequence.}
    \label{fig:refinement_terminal}
\end{figure}

The planetary tour beam search identified a high-scoring terminal sequence geometry involving repeated flybys of Planets 6 and 7. In a purely ballistic solution, this sequence could not be extended further. However, by using the solar sail and allowing the flyby times to vary, the terminal sequence could be refined to include additional flybys of these planets. This produced a significant increase in the final solution score.

\begin{table}[t]
\centering
\caption{Refined terminal sequence options.}
\begin{tabular}{lr}
\toprule
Terminal sequence & Unbonused weight \\
\midrule
$6$-$8$-$9$-$8$-$2$-$5$-$7$-$6$-$7$-$6$-$7$ & 128.928 \\
$6$-$8$-$9$-$8$-$2$-$5$-$7$-$6$-$7$-$6$-$7$-$6$ & 138.550 \\
$6$-$8$-$9$-$8$-$2$-$5$-$7$-$6$-$7$-$6$-$7$-$6$-$7$ & 151.891 \\
$6$-$8$-$9$-$8$-$2$-$5$-$7$-$6$-$7$-$6$-$7$-$6$-$7$-$6^{\ast}$ & 155.299 \\
\bottomrule
\end{tabular}
\label{tab:refined_terminal}
\end{table}

The refinement process is illustrated in Fig.~\ref{fig:refinement_terminal}, with the resulting sequence scores summarized in Table~\ref{tab:refined_terminal}. The process began by refining the terminal sequence identified in Section~\ref{sec:ballistic_tour} using \gls{scp}. This solve used a minimum initial start time of 100 years and allowed a free incoming flyby velocity at the start. The resulting sequence achieved an unbonused score of 128.928. Then, a further flyby was appended to the end of the sequence, initially placed at the maximum mission time of 200 years, and the full sequence was reoptimized. This increased the unbonused score to 138.550. Repeating the same procedure with an additional final flyby again produced a significant improvement.

It is important to note that the \gls{scp} refinement acts as a local optimizer and cannot substantially alter the underlying sequence geometry. For this reason, the flyby time at the Planet 2 \gls{ga} was also varied to test whether providing more time for the terminal sequence would improve the score. The best result was obtained by offsetting the Planet 2 encounter by two orbital periods later, producing what became the submitted terminal sequence with an unbonused score of 151.891.

Further analysis was then conducted to determine whether another Planet 6 flyby could be added to the end of the sequence. However, with the current solution geometry, this extension did not converge to a feasible solution. The best refinement required an impulsive correction of approximately 62 m/s at one of the Planet 7 \glspl{ga}. This indicates that the solution with the additional flyby was very close to feasible. If feasible, this would have increased the unbonused score to 155.299, as indicated by the starred entry in Table~\ref{tab:refined_terminal}.

A similar analysis was conducted for the start of the terminal phase. It was found that with the right timing, a further Planet 5 flyby could be added before the 8-9-8 triangular structure at the start of the terminal phase. This provided a small additional increase in the terminal sequence score and was included in the final assembled solution.

\subsection{Asteroid Belt Tours}

One option for obtaining the grand-tour bonus is to construct an asteroid tour using the solar sail within the asteroid belt. The core motivation was that if the asteroid geometry is favorable and the solar sail provides sufficient control authority, the required flybys may be achieved quickly despite the relatively low raw scientific weights of the asteroids.

To assess this possibility, a set of best-case asteroid tours was constructed using a modified beam search. The search is initialized from any asteroid, with a flyby velocity that produces a transfer to another object. These form the initial nodes. At each subsequent object, the same process as described in Section~\ref{sec:beam_search} is applied, but instead of requiring an exact flyby velocity match, a bounded mismatch is allowed. This enables the construction of sequences between massless bodies, where no \gls{ga} is available to change the spacecraft velocity. The search includes the asteroids, comets, and the dwarf planet Yandi.

The resulting sequences are not feasible in the ballistic model. However, the \gls{scp} refinement can remove the flyby velocity mismatches by using the solar sail and adjusting the flyby times. The objective of the \gls{scp} solve is then to minimize the total duration of the sequence, which would provide a best-case estimate of the time cost associated with completing an asteroid tour.

\begin{figure}[t]
    \centering
    \includegraphics[width=\textwidth]{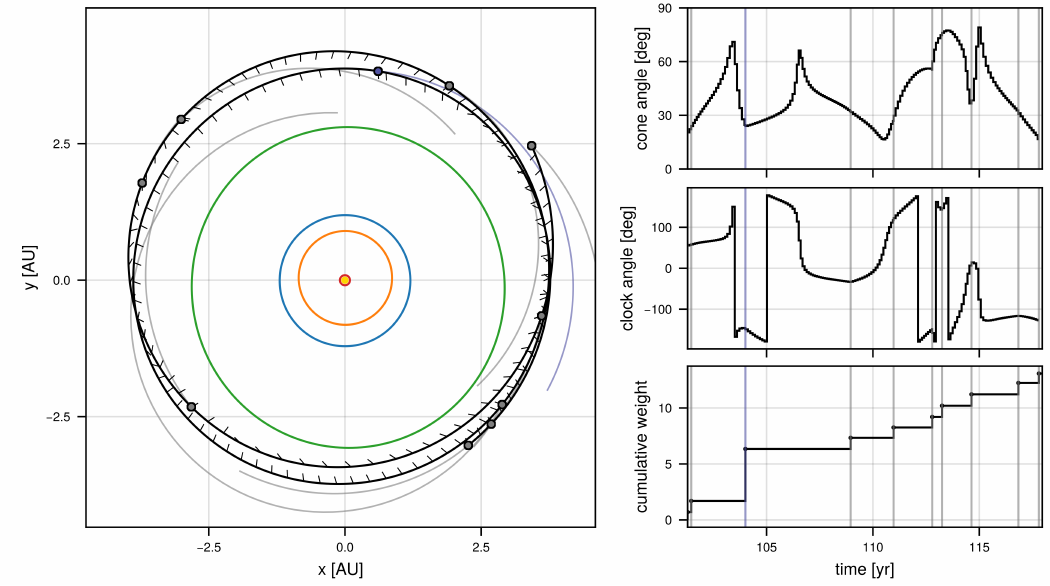}
    \caption{Solar sail refinement of a candidate asteroid belt tour.}
    \label{fig:asteroid_tour_refined}
\end{figure}

\begin{table}[t]
\centering
\caption{Improvement of an asteroid tour sequence using SCP.}
\begin{tabular}{lrrrr}
\toprule
ID & Predicted time & $V_\infty$ mismatch & Refined time & Refined \\
 & of flight [yr] &  [km/s] & of flight [yr] & weight \\
\midrule
1240 & --     & --    & --     & 0.702 \\
1239 & 0.173  & 0.386 & 0.171  & 0.997 \\
1000 & 2.540  & 0.303 & 2.546  & 4.647 \\
1007 & 4.941  & 0.343 & 4.947  & 0.993 \\
1148 & 2.031  & 0.022 & 2.018  & 0.930 \\
1133 & 1.813  & 0.067 & 1.816  & 0.938 \\
1254 & 0.460  & 0.015 & 0.460  & 1.006 \\
1257 & 1.387  & 0.044 & 1.385  & 1.021 \\
1240 & 2.201  & 0.613 & 2.205  & 1.011 \\
\midrule
Total & 15.546 & 1.794 & 15.548 & 13.086 \\
\bottomrule
\end{tabular}
\label{tab:asteroid_scp}
\end{table}

The best result from this analysis is shown in Fig.~\ref{fig:asteroid_tour_refined}, with the corresponding statistics given in Table~\ref{tab:asteroid_scp}. Before refinement, the total flyby-velocity mismatch across the sequence is 1.794 km/s. After \gls{scp} refinement, this mismatch is reduced to zero, while the total duration remains almost unchanged. Therefore, in the case of this sequence, the solar sail is able to correct the velocity mismatches without requiring a significant increase in time of flight.

This asteroid tour does not include the full set of 13 asteroids or comets required for the grand-tour bonus. Instead, it visits seven unique asteroids, with one repeated visit, as well as Yandi. The refined sequence scores 13.086 before global bonuses over 15.548 years, corresponding to an average of approximately 0.84 unbonused score per year. Much of this score is contributed by the Yandi flyby. This suggests that if a similar trajectory could be extended to the full 13 unique asteroid or comet flybys, an asteroid tour would require approximately 30 years in a best-case scenario. Since the grand-tour bonus is worth approximately 33 years in effective mission duration, such a tour would likely be beneficial, but only by a modest margin because the score rate during the asteroid tour is relatively low. As such, based on this analysis, the total score improvement enabled by a good-quality asteroid tour was estimated to be between 20 and 40.

\subsection{Refinement of Resonant Tours}
\label{sec:resonance_results}

\begin{figure}[t]
    \centering
    \includegraphics[width=\textwidth]{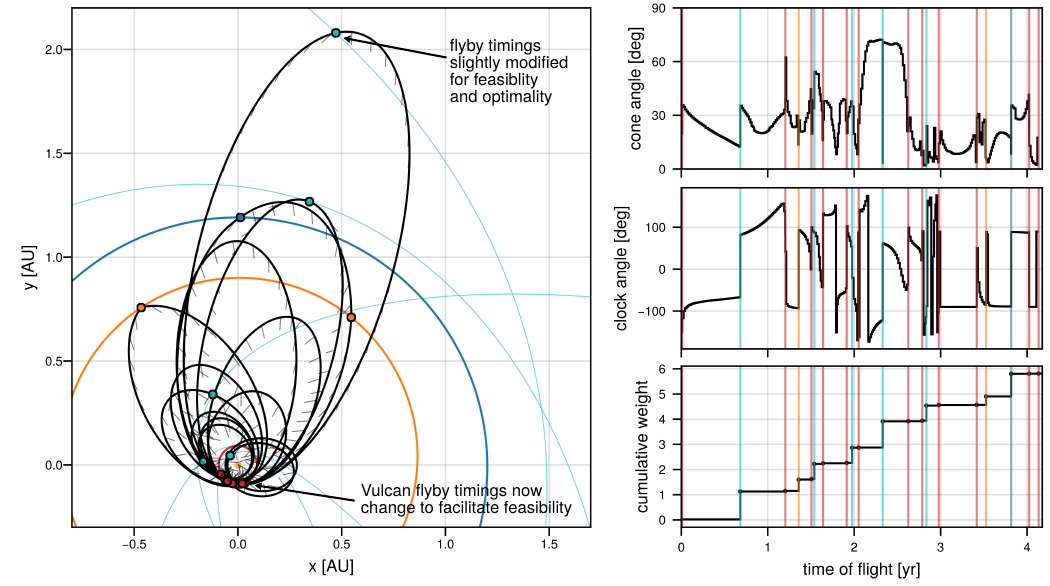}
    \caption{Example Vulcan-resonant tour solution refined using the solar sail SCP.}
    \label{fig:resonant_refined}
\end{figure}

Next, the \gls{scp} is used to refine the resonant Vulcan tours created through the beam search introduced in Section~\ref{sec:resonance}. The initial and final flyby times at Vulcan are fixed, and the incoming flyby velocity at Vulcan is fixed. All other flyby times and flyby velocities are free to change. The initial trajectory from the resonant beam search, shown in Fig.~\ref{fig:resonant_tour_beam} is used as the initial guess. The \gls{scp} then optimizes the solar sail control, the flyby timing, and the \gls{ga} geometry at Vulcan to produce a dynamically feasible trajectory that maximizes the total penalized scientific score of the trajectory.

An example of this refinement is shown in Fig.~\ref{fig:resonant_refined}. The most apparent change in this refinement process is that the Vulcan flyby times are no longer confined to the nominal single patch point. Instead, the \gls{scp} shifts these encounters over a wider range of times to satisfy the trajectory and \gls{ga} constraints.

In terms of the \gls{scp} formulation for this example problem, a solar sail control discretization of 5 days was used, resulting in a total of 329 nodes and 310 segments across 19 legs. The resulting optimization problem has a total of 7,803 variables and 5,503 constraints. This optimization completed in less than 10 seconds on a single thread of an AMD Ryzen 7 5800X3D processor, taking approximately 30 iterations for the \gls{scp} to converge.

The same refinement process was used in the submitted solution, with a much longer beam depth to cover the required duration. One final modification was made to slowly move the Vulcan match point around Vulcan's orbit throughout the resonant Vulcan phase. This allowed the search to sample a wider range of comet geometries. For example, a comet with a particular periapsis geometry relative to Vulcan may be better reached from one direction than another, and therefore may score more for some Vulcan phases. The resulting differences between consecutive departure and arrival Vulcan phases were small enough to be corrected by the \gls{scp}. This phase variation also helped reduce seasonal penalties by allowing comets to be visited at more diverse phases of their orbits.

The main limitation of this approach is that the beam search described in Section~\ref{sec:resonance} did not provide a reliable feasibility guarantee for the later \gls{scp} refinement. The pruning checks on the Vulcan flybys used within the beam search are only approximate, so some trajectories that pass the flyby-velocity and turning-angle filters cannot be made feasible by the \gls{scp}. For this reason, the beam search had to be run in a supervised manner in tandem with the \gls{scp}. After several beam-search expansions, the current best solution was refined using \gls{scp}. If the refinement did not converge, the move responsible for the infeasibility was identified and blocked explicitly in the beam search. The search was then backtracked and restarted. This supervised validation became the main bottleneck of the methodology, especially when attempting to push the solution quality by allowing larger flyby-velocity mismatch tolerances in the search.

\section{Solution Assembly}
\label{sec:assembly}

This section describes how the individual trajectory phases developed above were assembled into the submitted solution. Since each phase is constructed using a different beam search and/or refinement procedure, the phase boundaries must be selected and optimized carefully to produce a dynamically continuous trajectory.

\subsection{Joining Phases Together}

To produce a valid solution to \gls{gtoc13}, the individual phases must be joined to form a dynamically continuous trajectory. This process is important because the choice of connection point can introduce significant suboptimalities to the trajectory if it is made poorly. For example, if the resonant Vulcan phase ends in a state that cannot be connected efficiently to the terminal sequence, many flybys may need to be removed from the resonant phase to provide enough time for the trajectory to be patched.

For this reason, the exact choices of phase splitting were made carefully to preserve flexibility at each phase boundary. First, an appropriate entry sequence was selected; in this case, the 10-9-sun entry sequence was used. This sequence terminates after a close pass of Altaira, but does not by itself guarantee a useful connection to the following phase. Since the next phase is the resonant Vulcan phase, the 10-9-sun entry sequence was adapted to include a Vulcan flyby. The \gls{scp} was then applied with a minimum terminal flyby velocity objective. Since \gls{scp} is a local optimizer, this does not substantially change the Vulcan arrival time. Instead, it locally reduces the terminal flyby velocity with small changes to the Vulcan arrival time and the solar sail control. The result is a valid entry sequence with a fixed arrival time and a known incoming flyby velocity at Vulcan.

This Vulcan encounter then serves as the match point from which the resonant Vulcan beam search and refinement are initialized. In the \gls{scp} refinement of the resonant Vulcan phase, this is imposed through an initial \gls{ga} constraint, which then allows the initial outgoing flyby velocity at Vulcan to be optimized. The optimization of the resonant Vulcan phase then proceeds from this initial condition.

The desired end time of the resonant Vulcan phase was fixed by the start of the terminal sequence, which occurs at approximately 98 years. The beam search for the resonant phase was therefore run to produce sequences that terminate near this time. These sequences end at Vulcan with a specific incoming flyby velocity and flyby epoch, which could, in principle, be patched directly onto the terminal phase. However, this did not provide enough flexibility in practice. Instead, the patch was made one Vulcan \gls{ga} earlier, so that the final \gls{scp} connection problem included the sequence Vulcan-comet-Vulcan-(terminal phase). The initial time and incoming flyby velocity at the first Vulcan flyby were fixed, while the second Vulcan encounter was permitted to move. This provided enough timing flexibility at the second Vulcan flyby to connect the resonant phase to the terminal sequence without any loss in time.

\subsection{Scientific Flyby Selection}

\begin{figure}[b]
    \centering
    \includegraphics[width=\textwidth]{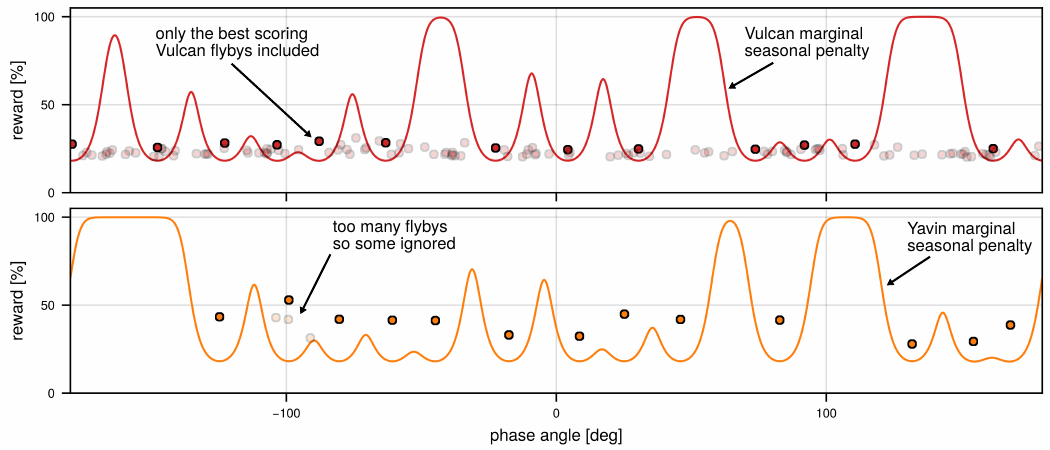}
    \caption{Selection results for scientific flybys including flyby-velocity and seasonal penalties.}
    \label{fig:scientific_flyby_selection}
\end{figure}

As the final stage of the solution process, a genetic algorithm was used to select which flybys should be marked as scientific before submission. The decision variables were represented by a boolean vector over the flybys that could optionally be made non-scientific, which in this solution were the Vulcan and Yavin flybys. Each boolean value then indicates whether the corresponding flyby is retained as scientific. The objective was to maximize the total scientific return after selecting the best 13 flybys. A genetic algorithm was used for this step because the selection problem has nonlinear interactions between scientific flybys that arise from the seasonal penalty.

The results of this process are shown in Fig.~\ref{fig:scientific_flyby_selection}. This illustrates the flyby-velocity penalty through the height of the scatter markers, together with the marginal seasonal penalty associated with adding each flyby to the scientific set. It is apparent that the selected flybys span a range of phase angles with each planet, which reduces the marginal seasonal penalty, while still favoring flybys with strong flyby-velocity penalty performance.

In retrospect, a better process would have been to select the scientific flybys first and then run the final \gls{scp} refinement with only those flybys contributing to the score. This may have allowed the final solution to be improved slightly. However, the expected improvement would be small, likely less than 0.5 in score. This is because only a small amount of score is lost through the flybys that were ultimately marked as non-scientific.

\section{Results}
\label{sec:results}
The solution submitted by \textit{TheAntipodes} placed third in the \gls{gtoc13} competition, demonstrating the effectiveness of the presented strategies. This section summarizes the final trajectory, its score breakdown, and the main features that contributed to the submitted result.

\subsection{Submitted Trajectory}

\begin{figure}[t]
    \centering
    \includegraphics[width=\textwidth]{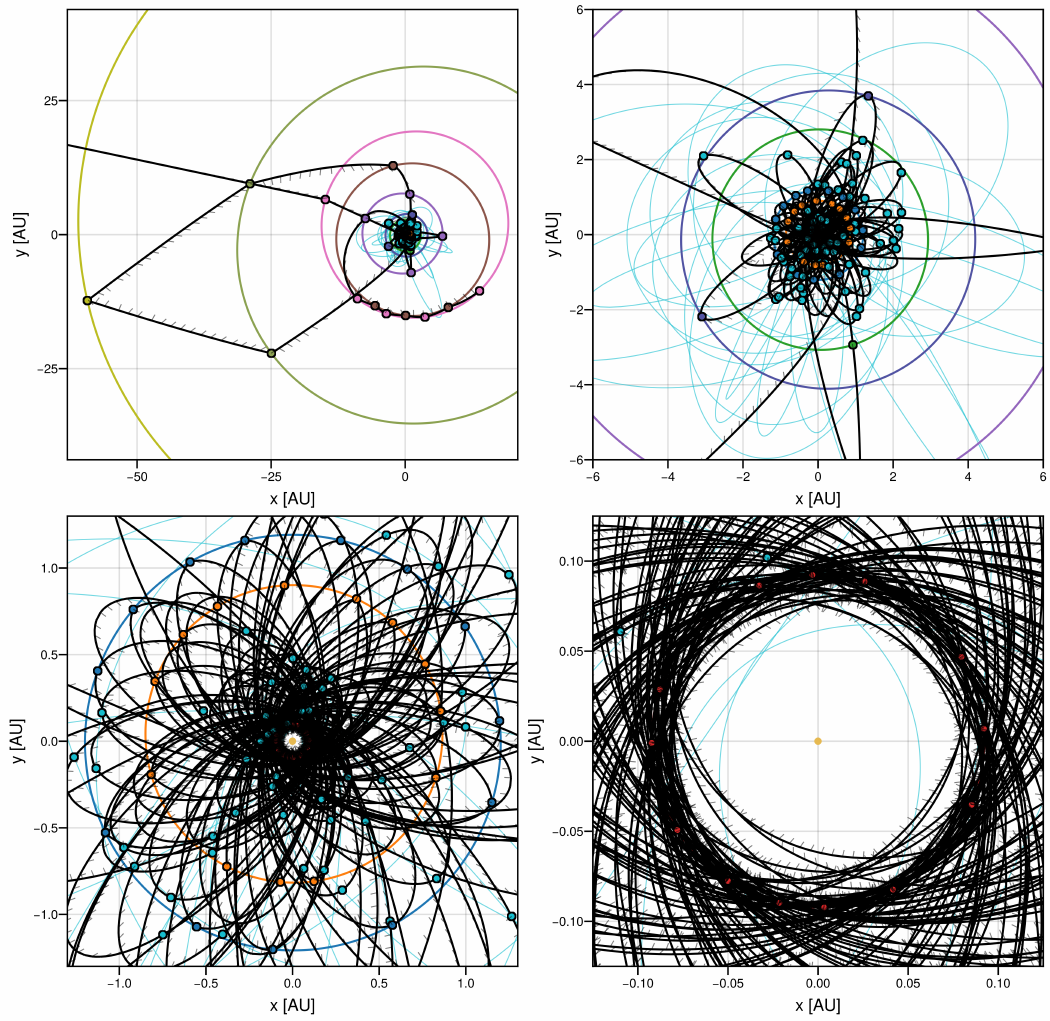}
    \caption{Final submitted solution to GTOC13 involving 133 scientific flybys for a total score of 337.878.}
    \label{fig:final_trajectory}
\end{figure}

\begin{figure}[t]
    \centering
    \includegraphics[width=\textwidth]{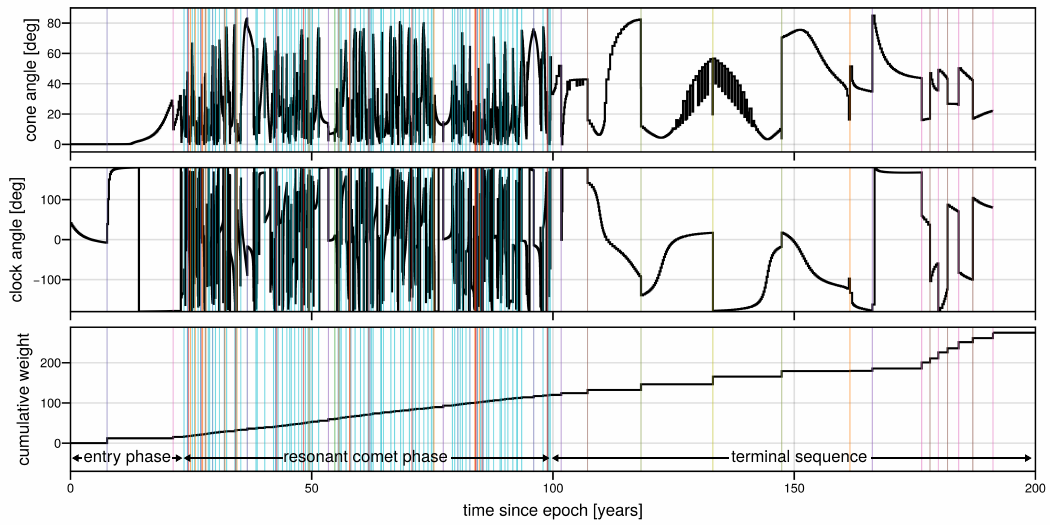}
    \caption{Final submitted solution details including solar sail cone and clock angle profiles.}
    \label{fig:final_trajectory_details}
\end{figure}

\begin{table}[b]
    \centering
    \caption{Final score breakdown for the submitted trajectory.}
    \begin{tabular}{lr}
        \toprule
        Metric & Value \\
        \midrule
        Raw science weight, all flybys & $557.900$ \\
        Non-scientific flyby weight & $-14.600$ \\
        \midrule
        Raw science weight, scoring flybys & $543.300$ \\
        Flyby-velocity penalty & $-265.079$ \\
        Seasonal penalty & $-3.778$ \\
        \midrule
        Penalized science score & $274.443$ \\
        Grand-tour bonus $\times 1.2$ & $+54.889$ \\
        Time bonus $\times 1.026$ & $+8.547$ \\
        \midrule
        Total score & $337.878$ \\
        \bottomrule
    \end{tabular}
    \label{tab:final-score-breakdown}
\end{table}

The submitted solution contains 252 flybys, of which 133 are selected as scientific, and achieves a final score of 337.878 after all penalties and bonuses are applied. The score breakdown is detailed in Table~\ref{tab:final-score-breakdown}. In total, the scientific flybys consist of 57 planetary flybys, two flybys of the dwarf planet Yandi, and 74 comet flybys. The solution does not include any asteroid flybys.

The overall geometry of the submitted trajectory can be seen at several different scales in Fig.~\ref{fig:final_trajectory}. The solution begins with the 10-7-sun entry sequence, with an initial velocity of 41.582 km/s. This entry phase scores approximately 0.676 unbonused score per year. It is followed by the resonant Vulcan phase, which lasts approximately 75 years and scores approximately 1.35 unbonused score per year. During this phase, most \glspl{ga} occur with the resonant target Vulcan, but several \gls{ga} with other planets are also included. These include the maximum allowed number of flybys of Planets 2 and 3, a single \gls{ga} of Planet 4, several flybys of Planet 5, and flybys of the dwarf planet Yandi. There are also a total of 74 comet flybys within this phase.

At a time of approximately 100 years, the trajectory leaves the Vulcan resonance and transfers into the terminal sequence. This begins with flybys of Planets 5 and 6, followed by the 8-9-8 triangular arrangement, a Planet 2 flyby, and finally a high-scoring sequence involving Planet 5 and repeated flybys of Planets 6 and 7. This final phase scores highly because the flybys occur at velocities that are not penalized by the flyby-velocity penalty. Considered on its own, the terminal sequence scores approximately 1.5 unbonused score per year, making it the highest score-rate phase of the submitted solution. The trajectory terminates at 191.177 yr, with no additional feasible flyby target available afterward. Since the trajectory visits all planets and at least 13 additional objects, it receives the grand tour bonus.

The solar sail control history is shown in Fig.~\ref{fig:final_trajectory_details}. The cone and clock angle profiles show the detailed solar sail control structure that is computed by the \gls{scp} optimizer. Some intervals exhibit chattering behavior, although this occurs primarily during phases where the solar sail has little effect on the trajectory, most notably near Planet 9. The special 0.01 AU periapsis limit is not used at any point in the submitted trajectory. The trajectory passes inside the orbit of Vulcan several times, but these close passages remain well outside the general periapsis constraint.

\subsection{Retrospective}

\begin{figure}[t]
    \centering
    \includegraphics[width=\textwidth]{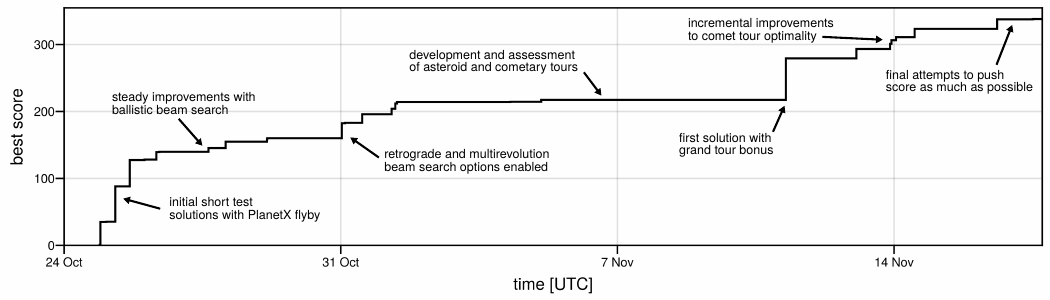}
    \caption{Evolution of \textit{TheAntipodes} solution score over the competition time frame.}
    \label{fig:solution-over-time}
\end{figure}

The \gls{gtoc13} problem presented a difficult challenge, but also admitted a wide range of successful solution strategies. \textit{TheAntipodes}' solution process evolved from ballistic beam searches, through attempts to construct asteroid-belt tours, and eventually to the construction of resonant Vulcan tours near Altaira. This progression over the competition time frame is shown in Fig.~\ref{fig:solution-over-time}, highlighting the incremental nature of the solution process. Later solutions built directly on the structures and insights obtained from earlier search phases.

After the competition, the quality of the resonant Vulcan tour phase appears to have been particularly important to the final score. In retrospect, this is the part of the solution approach that could have been improved the most. During the competition, the role of the \gls{scp} refinement in connecting the resonant flyby sequences was not fully understood. Initially, the solar sail was assumed to be directly modifying the flyby geometry at the Vulcan \gls{ga}; however, the main effect was instead that the sail allowed the Vulcan \gls{ga} times to shift slightly. Since Vulcan moves during this time shift, the incoming and outgoing flyby geometry also changes. A beam search strategy that accounted for this behavior and its limitations would likely have produced more feasible and better-quality resonant sequences, while reducing the need for the manual lockstep validation used during the resonant beam search.

After analyzing the solutions submitted by other teams, the terminal sequence presented in this work appears close to optimal and would be difficult to improve without a major change in sequencing. The largest remaining improvements in strategy are instead likely to come from the resonant phase. With better handling of this phase, such as that achieved by the first-place team, \textit{THU-LAD}, a maximum score of approximately 360--370 may have been possible using the final time bonus.

\section{Conclusions}

This paper presented the construction process behind the third-place solution to \gls{gtoc13}. The problem posed a challenging solar sail \gls{ga} tour-design task, requiring the simultaneous treatment of target sequencing, flyby timing, solar sail control, and final solution assembly. Several iterative solution strategies were developed and evaluated, with the final approach combining ballistic search, solar sail trajectory refinement, resonant tour construction, and targeted refinement of high-value tour segments.

Beam search provided an effective first tool for identifying ballistic \gls{ga} structures without using the solar sail. Although these trajectories were not expected to be competitive by themselves, they revealed favorable transfer geometries and helped guide later solar sail refinement. In particular, the decision to focus on a high-scoring terminal sequence, and then use this to define the connection target for a resonant Vulcan tour, proved central to the submitted solution.

More generally, \gls{scp} proved to be a powerful tool for the \gls{gtoc13} problem. It allowed many different solution structures to be evaluated and optimized efficiently without requiring high-performance computing resources. The combination of an adaptive-time formulation with a lossless convexification of the solar sail control constraints enabled rapid refinement and optimization of target sequences containing, in many cases, hundreds of \gls{ga}s.

\section*{Acknowledgments}

The authors thank Cristina Parigini and all other members of \textit{TheAntipodes} for their suggestions, tests, insights, and discussions throughout the competition.

\bibliographystyle{unsrtnat}
\bibliography{references}

\end{document}

%% file: acronyms.tex
\newacronym{gtoc13}{GTOC13}{13th Global Trajectory Optimization Competition}
\newacronym{gtoc}{GTOC}{Global Trajectory Optimization Competition}
\newacronym{scp}{SCP}{sequential convex programming}
\newacronym{srp}{SRP}{solar radiation pressure}
\newacronym{socp}{SOCP}{second-order cone programming}
\newacronym{dag}{DAG}{directed acyclic graph}
\newacronym{act}{ACT}{Advanced Concepts Team}
\newacronym{jpl}{JPL}{Jet Propulsion Laboratory}
\newacronym{minlp}{MINLP}{mixed-integer non-linear programming}
\newacronym{ga}{GA}{gravity assist}
\newacronym{ad}{AD}{automatic differentiation}
\newacronym{soc}{SOC}{second-order cone}
\newacronym{mip}{MIP}{mixed-integer programming}